\documentclass[acmsmall]{acmart}

\usepackage{standalone}

\usepackage{graphicx}

\usepackage{colortbl}
\usepackage{xcolor}
\usepackage{array}
\definecolor{light}{RGB}{222,235,247}
\definecolor{dark}{RGB}{158,202,225}
\definecolor{darker}{RGB}{49,130,189}
\newcolumntype{a}{>{\columncolor{light}}r}
\newcolumntype{b}{>{\columncolor{dark}}r}
\usepackage[ruled,lined,linesnumbered,vlined]{algorithm2e}

\SetCommentSty{mycommfont}

\usepackage{amsmath}
\usepackage{amsfonts}
\usepackage{amsopn}
\usepackage{amsthm}
\usepackage{nccmath} \usepackage{url}
\usepackage{multicol}
\usepackage{multirow}
\usepackage{setspace} \usepackage{xspace}
\usepackage{listings}
\usepackage{stmaryrd}
\usepackage{todonotes} \usepackage{booktabs}
\usepackage{threeparttable}
\usepackage[utf8x]{inputenc}
\usepackage[normalem]{ulem}	\usepackage{balance}
\usepackage{subcaption}
\usepackage[T1]{fontenc}

\usepackage[]{hyperref}

\usepackage[scaled=0.87]{beramono}

\usepackage[scaled]{helvet}
\usepackage{enumitem}
\usepackage{pdftexcmds}

\newcommand*{\proj}{\textsc{\textsf{Dflare}}\xspace}
\newcommand*{\repair}{\textsc{\textsf{Drepair}}\xspace}
\newcommand*{\projr}{$\text{\proj}_R$\xspace}
\newcommand*{\projd}{$\text{\proj}_D$\xspace}
\newcommand*{\NumOriginalCompressedParis}{21\xspace}

\newcommand*{\RepairRate}{48.48\%\xspace}

\makeatletter

\newcommand{\Rmnum}[1]{\expandafter\@slowromancap\romannumeral #1@}
\makeatother

\newcommand{\etal}{\hbox{\emph{et al.}}\xspace}

\newcommand{\eg}{\hbox{\emph{e.g.}}\xspace}
\newcommand{\ie}{\hbox{\emph{i.e.}}\xspace}
 
\newcommand{\wrt}{\hbox{\emph{w.r.t.}}\xspace}

\newcommand{\HighlightGrey}[1]{}
\newcommand{\HighlightBlack}[1]{}

\newcommand{\myparagraph}[1]{
\noindent \textit{\textbf{#1.}}\quad
}
\newcommand{\mymetircs}[2]{
\noindent \textit{\textbf{#1}} and \textit{\textbf{#2}}.\quad
}

\definecolor{codegreen}{rgb}{0,0.6,0}
\definecolor{codegray}{rgb}{0.5,0.5,0.5}
\definecolor{codepurple}{rgb}{0.58,0,0.82}
\definecolor{backcolour}{rgb}{0.95,0.95,0.92}
\lstdefinestyle{mystyle}{
commentstyle=\color{codegreen},
    keywordstyle=\color{magenta},
    numberstyle=\tiny\color{codegray},
    stringstyle=\color{codepurple},
    basicstyle=\ttfamily\footnotesize,
}
\lstdefinestyle{nobgstyle}{
    backgroundcolor=,
    commentstyle=\color{codegreen},
    keywordstyle=\color{magenta},
    numberstyle=\tiny\color{codegray},
    stringstyle=\color{codepurple},
    basicstyle=\ttfamily\footnotesize,
}

\SetKw{Continue}{continue}

\SetKwProg{Fn}{Function}{:}{}
\newcommand{\xmax}{x_\textit{max}}
\newcommand{\xt}{x_t}
\newcommand{\xs}{x_s}

\newcommand\ignore[1]{}

\usepackage{tikz}
\newcommand*\circled[1]{\tikz[baseline=(char.base)]{
		\node[shape=circle,draw,inner sep=1pt] (char) {#1};}}

\usepackage{cleveref}

\Crefname{table}{Table}{Tables}
\crefname{table}{Table}{Tables}
\Crefname{figure}{Figure}{Figures}
\crefname{figure}{Figure}{Figures}
\Crefname{algocf}{Algorithm}{Algorithms}
\crefname{algocf}{Algorithm}{Algorithms}
\Crefname{algorithm}{Algorithm}{Algorithms}
\crefname{algorithm}{Algorithm}{Algorithms}

\crefformat{chapter}{\S#2#1#3}
\crefmultiformat{chapter}{\S\S#2#1#3}{ and~#2#1#3}{, #2#1#3}{, and~#2#1#3}

\crefformat{section}{\S#2#1#3}
\crefmultiformat{section}{\S\S#2#1#3}{ and~#2#1#3}{, #2#1#3}{, and~#2#1#3}

\usepackage{tcolorbox}
\newcommand{\answer}[2]{ \begin{tcolorbox}
		\textbf{#1}: #2
\end{tcolorbox}}

\newcommand{\df}{\mbox{DiffChaser}\xspace}

\newcommand{\pool}{\texttt{pool}\xspace}

\AtBeginDocument{\providecommand\BibTeX{{\normalfont B\kern-0.5em{\scshape i\kern-0.25em b}\kern-0.8em\TeX}}}

\setcopyright{acmlicensed}
\acmJournal{TOSEM}
\acmYear{2023} \acmVolume{1} \acmNumber{1} \acmArticle{1} \acmMonth{1} \acmPrice{15.00}\acmDOI{10.1145/3583564}

\hypersetup{draft}

\begin{document}

\title{Finding Deviated Behaviors of the Compressed DNN Models for Image Classifications}

\titlenote{This is the author version. The DOI of the published version is \url{http://dx.doi.org/10.1145/3583564}}

\author{Yongqiang Tian}
\email{yongqiang.tian@uwaterloo.ca}
\orcid{0000-0003-1644-2965}
\affiliation{\institution{University of Waterloo}
	\city{Waterloo, ON}
	\country{Canada}}
\affiliation{
\institution{The Hong Kong University of Science and Technology}
\city{Kowlong, Hong Kong}
\country{China}}

\author{Wuqi Zhang}
\email{wzhangcb@cse.ust.hk}
\orcid{0000-0001-8039-0528}
\affiliation{\institution{The Hong Kong University of Science and Technology}
	\city{Kowlong, Hong Kong}
	\country{China}}

\author{Ming Wen}
\email{mwenaa@hust.edu.cn}
\orcid{0000-0001-5588-9618}
\affiliation{\institution{Huazhong University of Science and Technology}
	\city{Wuhan, Hubei}
	\country{China}}

\author{Shing-Chi Cheung}
\email{scc@cse.ust.hk}
\orcid{0000-0002-3508-7172}
\affiliation{\institution{The Hong Kong University of Science and Technology}
	\city{Kowlong, Hong Kong}
	\country{China}}

\author{Chengnian Sun}
\email{cnsun@uwaterloo.ca}
\orcid{0000-0002-0862-2491}
\affiliation{\institution{University of Waterloo}
	\city{Waterloo, ON}
	\country{Canada}}

\author{Shiqing Ma}
\email{shiqing.ma@rutgers.edu}
\orcid{0000-0003-1551-8948}
\affiliation{\institution{Rutgers University}
	\city{Piscataway, NJ}
	\country{USA}}

\author{Yu Jiang}
\email{jiangyu198964@126.com}
\orcid{0000-0003-0955-503X}
\affiliation{\institution{Tsinghua University}
	\city{Beijing}
	\country{China}}

\begin{abstract}
	Model compression can significantly reduce the sizes of deep neural network (DNN) models,
and thus facilitates the dissemination of sophisticated, sizable DNN models,
especially for their deployment on mobile or embedded devices.
However, the prediction results of compressed models may deviate from those of their original models.
To help developers thoroughly understand the impact of model compression, it is essential to test these models to find those \textit{deviated behaviors} before dissemination.
However, this is a non-trivial task because the architectures and gradients of compressed models are usually not available.

To this end, we propose \proj, a novel, search-based, black-box testing technique to
automatically find triggering inputs that result in deviated behaviors in image classification tasks.
\proj iteratively applies a series of mutation operations to a given seed image, until a triggering input is found.
For better efficacy and efficiency,
\proj models the search problem
as Markov Chains and leverages the Metropolis-Hasting algorithm
to guide the selection of mutation operators in each iteration.
Further, \proj utilizes a novel fitness function to
prioritize the mutated inputs that either cause
large differences between two models' outputs, or trigger previously unobserved models' probability vectors.
We evaluated \proj on \NumOriginalCompressedParis compressed models for image classification tasks with three datasets.
The results show that \proj not only constantly outperforms the baseline in terms of efficacy, but also significantly improves the efficiency:
\proj is 17.84x$\sim$446.06x as fast as the baseline in terms of time;
the number of queries required by \proj to find one triggering input is only 0.186\%$\sim$1.937\% of those issued by the baseline.
We also demonstrated that the triggering inputs found by \proj can be used to repair up to \RepairRate deviated behaviors in image classification tasks
and further decrease the effectiveness of \proj on the repaired models. \end{abstract}



\maketitle

\section{Introduction}

\begin{figure}[h]

	\begin{subfigure}[b]{.45\linewidth}
		\centering
		\includegraphics[width=0.5\linewidth]{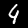}
	\end{subfigure}
	\hfil
	\begin{subfigure}[b]{.45\linewidth}
		\centering
		\includegraphics[width=0.5\linewidth]{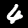}
	\end{subfigure}
	\hfil
	\caption{Images triggering deviated behaviors between LeNet-4 and its quantized model.
	The ground truth labels of both images are ``4'' and both of them are correctly classified as ``4'' by the original model. However, the quantized model classifies them as ``9'' and ``6'' respectively.}
	\label{fig:inconsistency_example}
\end{figure}

Compressing DNN models is one \emph{critical} stage in model dissemination,
especially for deploying sizable models on mobile or embedded devices
with limited computing resources.
Compared to their original models, compressed ones achieve similar prediction accuracy
while requiring significantly less time, processing power, memory and energy, for inference~\cite{compressionsurvery, HAQ}.
However, model compression is a lossy process:
given the same input, a compressed model can make predictions deviated from its original model ~\cite{diffchaser, deephunter}.
For example, given the two images in~\cref{fig:inconsistency_example}, the LeNet-4~\cite{Lecun98gradient-basedlearning} model correctly predicts both images
as \texttt{4} while its compressed model predicts the left one as \texttt{9} and the right one as \texttt{6}. We say that a deviated behavior occurs if a compressed model makes a prediction different from the one of the original model.
The input that triggers such a deviated behavior is referred to as a \textit{triggering input}.
Our objective is to find the triggering inputs for a given pair of a compressed model and the original one, so that the compressed model's quality can be further assessed before its dissemination beyond the dataset
that is used during model compression~\cite{8294186, 10.5555/3201607.3201772}.

It is preferred to find triggering inputs quickly so that developers can obtain in-time feedback to assess and
facilitate the entire dissemination workflow.
However, this is a challenging task.
Specifically,
to accelerate the inference speed and reduce storage consumption,
compressed models usually
do not expose their architectures or intermediate computation results via APIs~\cite{compressionsurvery}.
Gradient, one of the most common information leveraged by previous test generation approaches~\cite{deeptest,deepxplore,deephunter}, is also not always available in compressed models, especially for integer weights (See~\cref{subsec:assumption} for more details).
Without such information as guidance, it is difficult
for input generation techniques to efficiently find the triggering inputs.
For example, the state of the art, \df, requires thousands of queries from a pair of models to find a triggering input.
Considering the fact that the datasets in deep learning applications usually consist of more than thousands of inputs, such an inefficient approach not only incurs unaffordable computation workload to developers, but also compromises its practicality in industry.

In this paper, we propose \proj, an
effective and efficient
technique to automatically find triggering inputs for compressed DNN models that are designed for image classification tasks.
Given a non-triggering input as a seed, \proj mutates the seed continuously until a triggering input is found.
The mutation is guided by a specially designed fitness function, which measures
(1) the difference between the prediction outputs of the original and compressed models, and
(2) whether the input triggers previously unobserved probability vectors of the two models. The fitness function of \proj does not require the model's intermediate results, and thus \proj
is general and can be applied to any compressed model for image classifications.
Unlike \df~\cite{diffchaser}, \proj only selects one mutation operator and generates one mutated input at each iteration, resulting in much fewer queries than \df.
As another key contribution,
\proj models the selection of mutation operators as a Markov Chain process and adopts the Metropolis-Hastings (MH) algorithm~\cite{mcmc} to guide the selection.
Specifically, \proj prefers a mutation operator that is likely to increase the fitness function value of subsequent mutated input in the future.

To evaluate \proj, we construct a benchmark consisting of \NumOriginalCompressedParis pairs of models (\ie, each pair includes the original model and the corresponding compressed one) on three commonly used  image classification datasets: MNIST~\cite{lecun-mnisthandwrittendigit-2010}, CIFAR-10~\cite{cifar10} and ImageNet~\cite{imagenet_cvpr09}.
The compressed models are generated with diverse, state-of-the-art techniques:
weight pruning, quantization and knowledge distillation.
The model architectures include both small- and large-scale ones, from LeNet to VGG-16.

We evaluate \proj \wrt its effectiveness and efficiency and compare it with \df, the state-of-the-art black-box approach.
For effectiveness, we feed a fixed number of seed inputs to \proj and measure the ratio of seed inputs for which \proj can successfully generate triggering inputs.
For efficiency, we measure the time and queries that \proj needs to find one triggering input given a seed input.
The results show that \proj constantly achieves 100\% success rate while the baseline \df fails to do so, whose success rate drops to <90\% for certain cases in CIFAR,
and drops to around 20\% in ImageNet dataset.
More importantly, \proj can significantly improve efficiency.
On average, \proj can find a triggering input with only 0.52s and 24.99 queries, while \df needs more than 52.23s and 3642.50 queries.
In other words, the time and queries needed by \proj are only 0.99\% and 0.69\% of \df, respectively.

We conduct a case study
to further demonstrate the usefulness of \proj in model dissemination.
Specifically, we demonstrate that given a set of compressed models whose accuracy is very close to each other,
\proj can efficiently provide extra information to approximate the likelihood that the compressed model behaves differently from the original one.
Such in-time information can provide developers with more comprehensive evaluations towards compressed models, thus facilitating the selection of compressed models and the compression configurations in the dissemination of image classification models.

We explore the possibility to repair the deviated behaviors using the
triggering inputs found by \proj.
Our intuition is that the substantial amount of triggering inputs found by \proj
contains essential characteristics of such triggering inputs,
and may thus be used to train a separate repair model to fix the deviated behaviors of compressed models found by \proj for image classifications.
We design a prototype named \repair, serving as a post-processing stage of compressed models.
After the compressed model outputs the probability vector of
an arbitrary input,
\repair takes this vector as input and
aims to generate the same label as the one outputted by the original model.
We build \repair based on Single-layer Perceptron~\cite{slp} and train it using
the triggering inputs found by \proj and seed inputs.
Our evaluation shows that \repair can reduce up to \RepairRate deviated behaviors
and decrease the effectiveness of \proj on the repaired models.

\noindent\textit{\textbf{Contributions.}}
Our paper makes the following contributions.
\begin{enumerate}[topsep=5pt]
	\item We propose \proj, a novel, search-based,
	      guided testing technique
	      to find triggering inputs for compressed models for image classifications, to
	      help analyze and evaluate the impact of model compression.

	\item
	      Our comprehensive evaluations on a benchmark
	      consisting of \NumOriginalCompressedParis pairs of original and compressed
          image classification models
	      in diverse architectures
	      demonstrate that \proj significantly outperforms the state of
	      the art in terms of both effectiveness and efficiency.

   \item
   		  We demonstrated that the triggering inputs found by \proj
   		  can be used to repair up to \RepairRate deviated behaviors in image classification tasks
   		  and decrease the effectiveness of \proj on the repaired models.

	\item
	      To benefit future research,
	      we have made our source code and benchmark publicly available for reproducibility at \url{https://github.com/yqtianust/DFlare}
\end{enumerate}
 \section{Preliminary}
\label{sec:preliminary}

In this section,
we first introduce our scope and give a brief introduction about model compression.
Second, we present the annotations and assumptions used in this study
and the state-of-the-art technique.
At last, we discuss
the difference
between triggering inputs and adversarial samples.

\subsection{Scope}
\label{sec:scope}
Our technique focuses on the compressed DNN models for image classifications.
Image classification is one of the most important applications of deep learning and DNN compression techniques.
There are enormous studies in model compression
focusing on deploying compressed image classification models
resource-constrained device,
such as~\cite{7780890,
    deepcompression, compressionsurvery,9384353,10.1145/3489517.3530400,
    DBLP:conf/iclr/CaiGWZH20,9008764,DBLP:conf/iclr/ZhuG18,10.5555/3367471.3367522,DBLP:conf/iclr/LinSBDJ20,
    9794700}.
The deployment of compressed models for image classifications is also paid close attention by industries.
Mobile hardware vendors, such as Arm,\footnote{\url{https://developer.arm.com/documentation/102561/2111}} Qualcomm,\footnote{\url{https://developer.qualcomm.com/project/image-classification-qcs610-development-kit}}
and NVIDIA\footnote{\url{https://docs.nvidia.com/metropolis/TLT/tlt-user-guide/text/image_classification.html}}
provide detailed documentation to deploy image classification on their mobile devices.
Moreover, there are plenty of publicly available original models and compressed models for our research~\cite{pytorchmodels,nzmora2019distiller}
and their detailed instructions allow us to faithfully reproduce their results.
Moreover, many previous testing studies for DNN models also
primarily focus on image classification tasks~\cite{deephunter, deepmutation, cw,fgsm, DBLP:conf/icml/ZhangYJXGJ19}.
The baseline~\cite{diffchaser} used in our evaluation also concentrates on
the compressed models for image classifications.

Our study aims to find the triggering inputs that are \textit{not} in the original training set
or test set.
The model compression techniques are designed to compress the original model while preserving the accuracy as much as possible~\cite{compressionsurvery}.
As a result, the number of triggering inputs in
the training set and test set for the original model are pretty limited.
If there were a significant number of triggering inputs in the original training set and test set,
compressed models are likely to have a clear difference from the original models in terms of accuracy.
Developers can easily notice such triggering inputs by inspecting the accuracy
and then strive to fix the problematic compression processing
before deploying these models.
However, the triggering inputs outside the original datasets are not directly available to developers.
Finding them can help developers comprehensively evaluate their compressed models before the deployment.

\cref{tab:prevalence} lists the number of triggering inputs in the training set and test set
for three pairs of models used by \df.
The triggering inputs in the training set imply that such deviated behaviors may be related to the inherent proneness of model compression to deviating compressed models from their original models.
However, the number of triggering inputs in the training and test set is negligible ($\leq 0.62\%$).
These results may mislead the developers of compressed models, \eg,
developers may believe that the compressed models have almost identical behaviors
as their original models.
However, as shown by \df~\cite{diffchaser} and later in our evaluation,
there are a significant number of triggering inputs that are not in
the training set
or test set.
These extra triggering inputs can help developers comprehensively
evaluate their compressed models and repair the deviated behaviors.

\begin{table}[h]
    \centering
    {
        \caption{The numbers of triggering inputs and their percentages in training and test set.}
        \label{tab:prevalence}
       \begin{tabular}{@{}cccrr@{}}
                \toprule
                \textbf{Dataset}           & \textbf{Original Model} & \textbf{Compression Method} & \textbf{Training set}  & \textbf{Test set}\\ \midrule \midrule
                \multirow{2}{*}{MNIST}   & LeNet-1                 & Quantization-8-bit & 83 / 60000  = 0.13\% & 9 / 10000 = 0.05\%     \\ \cmidrule(l){2-5}
                & LeNet-5                 & Quantization-8-bit &  23 / 60000  = 0.38\%  & 5 / 10000 = 0.05\% \\ \midrule
                CIFAR-10                 & ResNet-20                 & Quantization-8-bit & 75 / 50000 = 0.15\%& 62 / 10000 = 0.62\% \\ \midrule
        \end{tabular}}
\end{table}
\subsection{Model Compression}
Model compression has become a promising research direction to
facilitate the deployment of deep learning models~\cite{10.1007/s10462-020-09816-7, compressionsurvery}.
The objective of model compression is to compress the large model into compact models
so that the compressed models are able to be deployed in resource-constrained devices,
such as the Internet of Things (IoT) and mobile phones.
Various model compression techniques have been proposed to reduce the size of DNN models
and the majority of them can be classified into the following three categories.

\myparagraph{Pruning}
Pruning is an effective compression technique to reduce
the number of parameters in DNN models~\cite{pruning,pruning2}.
Researchers find that considerable parameters in DNN models have limited
contribution to inference results~\cite{10.1007/s10462-020-09816-7,pruning,pruning2}
and removing them does not significantly decrease the model performance on the original test sets.
Pruning techniques can be further classified into several categories,
according to the subjects to be pruned,
including weights, neurons, filters and layers.
Weight pruning zeros out the weights of the connections between neurons
if the weights are smaller than some predefined threshold.
Neuron pruning removes neurons and their incoming and outgoing connections
if their contribution to the final inference is negligible.
In filter pruning, filters in convolutional layers
are ranked by their importance according to their influence on the prediction error.
Those least important filters are removed from the DNN models.
Similarly, some unimportant layers can also be pruned
to reduce the computation complexity of the models.

\myparagraph{Quantization}
Quantization compresses a DNN model by changing the number of bits
to represent weights~\cite{quantization,quantization2}.
In DNN models, weights are usually stored as 32-bit floating-point numbers,
After quantizing these weights into 8-bit or 4-bit,
the size of models can be significantly reduced.
Meanwhile, the quantized models consume less memory bandwidth than the original models.
A recent research direction of quantization is
Binarization~\cite{10.1609/aaai.v33i01.33013854, electronics8060661}.
It uses 1-bit binary values to represent the parameters of DNN models and
the model after binarization is referred to as Binarized Neural Networks (BNNs).

\myparagraph{Knowledge Distillation}
Knowledge distillation transfers the knowledge learned by original DNN models
(referred to as teacher models)
to compact models (\ie, student models)~\cite{10.1145/1150402.1150464, distillation, distillation2}.
After teacher models are properly trained using training sets,
student models are trained to mimic the teacher models.
We refer interested readers to a recent literature review~\cite{kdsurvey} for details.

\subsection{Annotations}
Let $n$ be the number of all possible classification
labels in a single-label image classification problem,
\ie, an image is expected to be correctly classified into only one of the $n$ labels.
Let $f$ denote a DNN model designed for this single-label image classification,
and $g$ denote a corresponding compressed model.
Given an arbitrary image as input $x$, model $f$ outputs a probability vector $f(x) = [p_1, p_2, p_3, \cdots, p_n]$.
We refer to the highest probability in $f(x)$ as \textit{top-1 probability} and denote  it  as $p_{f(x)}$.
We refer to the label whose probability is $p_{f(x)}$ in $f(x)$ as \textit{top-1 label} and denote it as $l_{f(x)}$.
Similarly, the probability vector of the compressed model, the top-1 probability and its label are denoted as $g(x) = [p_1^{\prime}, p_2^{\prime}, p_3^{\prime}, \cdots,  p_n^{\prime}]$,  $p_{g(x)}$ and $l_{g(x)}$, respectively.

\subsection{Assumptions}
\label{subsec:assumption}
We assume that the compressed model $g$ is a black-box
and only the information $g(x)$, $p_{g(x)}$ and
$l_{g(x)}$ are available~\cite{simpleattack,DBLP:conf/iclr/ChengLCZYH19,DBLP:conf/eccv/BhagojiHLS18,DBLP:conf/cvpr/ShiWH19}.
The internal states of models, including intermediate computation results,
neural coverages and gradients, are not accessible.
We make this assumption for the following reasons.

	First, in practice, the intermediate results of compressed models, such as activation values and gradients, are not available due to the lack of appropriate API support in deep learning frameworks.
	Modern deep learning frameworks, such as TensorFlow Lite~\cite{tflite} and ONNX Inference~\cite{onnx},
    usually provide APIs only for end-to-end inference of the compressed model,
    but not for querying intermediate results.
	The design decision of
    discarding intermediate results is mainly to improve inference efficiency~\cite{compressionsurvery,onnx}.

	Second, gradient information is not generally meaningful for some compressed models.
	For example, for the model that uses integer weights, their gradients are not defined and thus cannot be acquired.
	\cref{ex:code} shows such an example.
	The code snippet in \cref{ex:float-code} computes the gradient of $y=x^3$ with respect to the float tensor $x$ and running this code correctly outputs the expected gradient, \ie 12.
	The code in \cref{ex:int-code} also computes the gradient $y=x^3$ with respect to $x$, but the tensor $x$ in \cref{ex:int-code} is an integer tensor.
	Executing the code in \cref{ex:int-code} leads to a runtime error shown in \cref{ex:err-msg}.

	Third, if the compressed model under test requires special devices such as mobile phones, or the model is compressed on the fly, such as TensorRT~\cite{tensorrt}, accessing the intermediate results requires support from system vendors, which is not always feasible.
	The assumption of treating compressed models as a black-box increases the generalizability of \proj.

\begin{figure}
\begin{subfigure}[t]{0.425\linewidth}
	\begin{lstlisting}[language=python, frame=single,
showstringspaces=false]
import torch
x = torch.tensor([2.])
x.requires_grad=True
y = x**3
y.backward()
print(x.grad)
	\end{lstlisting}
	\caption{A code snippet to compute the gradient of \texttt{y} \wrt float tensor \texttt{x}.}
	\label{ex:float-code}
\end{subfigure}
\hfil
\begin{subfigure}[t]{0.425\linewidth}
	\begin{lstlisting}[
		language=python, frame=single,
		showstringspaces=false]
import torch
x = torch.tensor([2.]).int()
x.requires_grad=True
y = x**3
y.backward()
print(x.grad)
\end{lstlisting}
\caption{A code snippet to compute the gradient of \texttt{y} \wrt integer tensor \texttt{x}.}
\label{ex:int-code}
\end{subfigure}
\begin{subfigure}{0.9\linewidth}
\begin{lstlisting}[	language=bash, frame=single,
	showstringspaces=false]
Traceback (most recent call last):
	File "int_gradient.py", line 3, in <module>
		x.requires_grad=True
RuntimeError: only Tensors of floating point and complex dtype can require gradients
\end{lstlisting}
\caption{Error message when executing the code in (\subref{ex:int-code}).}
\label{ex:err-msg}
\end{subfigure}

\caption{Example code snippet of computing gradient for tensor with float weight (\cref{ex:float-code}) and integer weight (\cref{ex:int-code}) in PyTorch, respectively.
Executing the code in \cref{ex:float-code} outputs the correct value, \ie 12, while executing the code in \cref{ex:int-code} throws runtime error in \cref{ex:err-msg}.
}
\label{ex:code}
\end{figure}

\subsection{State of the Art}
\label{subsec:sota}
\df~\cite{diffchaser} is a black-box genetic-based approach to finding triggering inputs for compressed models.
In the beginning, it creates a pool of inputs by mutating a given non-triggering input.
In each iteration, \df crossovers two branches of the selected inputs and then selectively feeds them back to the pool until any triggering input is found.
To determine whether each mutated input will be fed back to the pool,
\df proposes \textit{k-Uncertainty} fitness function
and uses it to measure the difference between the highest probability and k-highest probability of \textit{either} $f(x)$ or $g(x)$.
Please note that \textit{k-Uncertainty} does not capture the difference between two models, resulting in its ineffectiveness in certain cases, as shown later in \cref{sec:eval}.
Another limitation is that the genetic algorithm used in \df needs to crossover a considerably large ratio of inputs and feed them into DNN models in each iteration.
As a result, it requires thousands of queries from the two models to find a triggering input.
Such a large amount of queries incur expensive computational resources, which are generally unavailable for devices that have limited computation capabilities, such as mobile phones and Internet-of-Things (IoT) devices.

There are many white-box test generation approaches for a single DNN model~\cite{surprise, deeptest, deepgauge, deephunter, deepxplore}.
However, they all need to access the intermediate results or gradient to guide their test input generation.
Thus, it is impractical to adopt them to address the
research problem of this paper.

\subsection{Differences from Adversarial Samples}
Adversarial samples are different from triggering inputs.
The adversarial attack approach targets a \emph{single} model using a malicious input, which is crafted by applying human-imperceptible perturbation on a benign input~\cite{cw, fgsm, tensorfuzz, deepxplore,deepsearch}.
In contrast, a triggering input is the one that can cause an inconsistent prediction between \emph{two} models,
\ie, the original model and its compressed model.
Note that adversarial samples of the original model are often not triggering inputs for compressed models.
In our preliminary exploration, we have leveraged FGSM~\cite{fgsm} and CW~\cite{cw} to generate adversarial samples for three compressed models using MNIST.
On average, only 18.6 out of 10,000 adversarial samples are triggering inputs.
Recent studies pointed out that compressed models can be an effective approach
to defend against adversarial samples~\cite{chen2021adversarial, 8854377}.

 \section{Methodology}
This section formulates the targeted problem, and then
details how we tackle this problem in \proj.

\subsection{Problem Formulation}
Given a non-triggering input as seed input $\xs$,
\proj strives to find a new input $\xt$ such that the
\mbox{top-1} label $l_{f(x_t)}$ predicted by the original model $f$ is different from the top-1 label $l_{g(x_t)}$ from the compressed model $g$, \ie, $l_{f(x_t)} \neq l_{g(x_t)}$.
Similar to the mutated-based test generations~\cite{diffchaser,tensorfuzz},
\proj attempts to find $\xt$ by applying a series of input mutation operators on the seed input $\xs$.
Conceptually, $\xt = \xs + \epsilon$, where $\epsilon$ is a perturbation made by the applied input mutation operators.

\subsection{Overview of \proj}

\begin{algorithm}[h]

	\KwIn{$\xs$: a seed input}
	\KwIn{$f$: the original model}
	\KwIn{$g$: the compressed model}
	\KwIn{\pool: a list of predefined input mutation operators}
	\KwIn{$timeout$: the time limit for finding a triggering input}
	\KwOut{an triggering input \textbf{$x_t$}}

	$op \gets $ an operator randomly selected from \pool \label{alg:line:select-first-op}\\
	$\xmax \gets \xs$ \label{alg:line:max-init} \\
	\Repeat{timeout}{
		$x \gets op(\xmax)$ \label{alg:line:mut} \\
		\If {$l_{f(x)} \neq l_{g(x)}$ \label{alg:line:checktrigger}}{
			$x_t \gets x$	\tcp{$x_t$ is a triggering input}
			\textbf{return} $x_t$ \quad
		}
		\If{$H_{f, g}(x) \geq H_{f,g}(\xmax) $ \label{alg:line:checkfit}}{
			$\xmax \gets x$ 	\label{alg:line:changex} \quad \tcp{if it has higher fitness value}
			$op$.update() \label{alg:line:increase} \quad \tcp{update its ranking value}
		}
		$op \gets pool$.select($op$)	\quad \tcp{select the next operator} \label{alg:line:selectop}
	}

	\caption{Overview of \proj}
	\label{alg:main}
\end{algorithm}
\cref{alg:main} shows the overview of \proj.
\proj takes four inputs: a seed input $\xs$, the original model $f$, the compressed model $g$,
and a list \pool of predefined input mutation operators;
it returns a triggering input $\xt$ if found.

\proj finds $\xt$ via multiple iterations.
Throughout all iterations, \proj maintains two variables:
$op$ is the input mutation operator to apply, which is initially randomly picked from \pool on~\cref{alg:line:select-first-op}
and updated each iteration on \cref{alg:line:selectop};
$\xmax$ is the input with the maximum fitness value
among all generated inputs, which is initialized with $\xs$ on~\cref{alg:line:max-init}.

In each iteration,
\proj applies an input mutation operator on the input which has the highest fitness value
to generate a new, mutated input, \ie, $x \gets op(\xmax)$ on~\cref{alg:line:mut}.
If $x$ triggers a deviated behavior between $f$ and $g$ on~\cref{alg:line:checktrigger},
then $x$ is returned as the triggering input $\xt$.
Otherwise, \proj compares the fitness values of $x$ and $\xmax$ on \cref{alg:line:checkfit},
and use the one that has the higher value for the next iteration (\cref{alg:line:changex}) .
The mutation operators are implemented separately from the main logic of \proj, and it is easy to integrate more mutation operators.
In our implementation, we used the same operators as \df.

Two factors can significantly affect the performance of \cref{alg:main}:
\emph{fitness function} and \emph{the strategy to select mutation operators},
of which both are detailed in the remainder of this section.

\subsection{Fitness Function}
Following the existing test generation approaches in software testing~\cite{jvmdiff,tensorfuzz,diffchaser}, in \proj, if the mutated input $x$ is a non-triggering input, the fitness function $H_{f,g}$ is used to determine whether $x$ should be used in the subsequent iterations of mutation (\cref{alg:main}, line~\ref{alg:line:checkfit}$\sim$\ref{alg:line:changex}).
By selecting the proper mutated input in each iteration, we aim to move increasingly close to the triggering input from the initial seed input $\xs$.

\subsubsection{Intuitions of \proj}
We design the fitness function from two perspectives.
First, if $x$ can cause a larger distance between
outputs of $f$ and $g$ than $\xmax$, $x$ is more favored than $\xmax$.
The intuition is that if $x$ can,
then future inputs generated by mutating $x$ are more likely to further enlarge the difference.
Eventually, one input generated in the future will increase the distance substantially
such that the labels predicted by $f$ and $g$ become different,
and this input is a triggering input that \proj has been searching for.

Second, when $x$ and $\xmax$ cause the same distance between outputs of $f$ and $g$,
 $x$ is preferred over $\xmax$
if $x$ triggers a
	previously unobserved model state in $f$ or $g$.
Conceptually, a model state refers to the internal status of the original or compressed models
during inference, including but not limited to a model's activation status.
If an input $x$ triggers a model state that is different from the previously observed ones, it is likely that it triggers a new logic flow in $f$ or $g$.
By selecting such input for next iterations, we are encouraging \proj to explore more new logic flows of two models, resulting in new model behaviors, even deviated ones.
Since the internal status of compressed models is not easy to collect, we use the probability vector to approximate the model state.

\subsubsection{Definition of Fitness Function}
\label{sec:fitness}
Now we present the formal definition of our fitness function $H_{f, g}(x)$ for a non-triggering input $x$
as a combination of two intuitions.

For the first intuition, given an input $x$, we denote the distance between two DNN models' outputs as
$\mathcal{D}_{f,g}(x)$.
Since $x$ is a non-triggering input, the top-1 labels of $f(x)$ and $g(x)$ are the same and we simply use the top-1 probability to measure the distance, \ie, $$\mathcal{D}_{f,g}(x) = |p_{f(x)} - p_{g(x)}| \in \left[0,1\right)$$

For our second intuition,
we use the probability vector to approximate the model state.
When executing \cref{alg:main},
we track the probability vectors produced by $f$ and $g$ on all generated inputs.
In the calculation of fitness value of $x$ at each iteration,
we check whether the pair of probability vectors output by the two DNN models $\left(f(x), g(x)\right)$
is observed previously or not.
Specifically, we adopt the Nearest Neighborhood algorithm~\cite{flann} to determine $\mathcal{O}(x)$, \ie, whether $\left(f(x), g(x)\right)$ is close to any previously observed states.
The result is denoted as $\mathcal{O}(x)$,
\[
\mathcal{O}(x) = \begin{cases}
  1 & \text{if $\left(f(x), g(x)\right)$ has \textit{not} been observed}  \\
  0 & \text{otherwise}
\end{cases}
\]

\noindent The fitness function $H_{f, g}(x)$ for a non-triggering input $x$  is defined as:
\[
H_{f, g}(x) = \delta^{-1}*\mathcal{D}_{f,g}(x) + \mathcal{O}(x)
\]
Specifically, according to $H_{f, g}$, for two non-triggering inputs,
	we choose the one with a higher $\mathcal{D}_{f,g}$ component.
If their $\mathcal{D}_{f,g}$ components are very close (\ie, the difference is less than the tolerance $\delta$), they will be chosen based on $\mathcal{O}(x)$.
In our implementation, we set $\delta=1\mathrm{e}{-3}$.

\subsection{Selection Strategy of Mutation Operators}
Existing work on test generation for traditional software has shown that the selection strategy of mutation operators can
have a significant impact on the performance of mutation-based test generation techniques adopted by \proj~\cite{Athena,jvmdiff}.
Following prior work, in each iteration, \proj favors a mutation operator
that has a high probability to make the next mutated input $x$ have a higher fitness value than $\xmax$.
Unfortunately, it is non-trivial to obtain such prior probabilities of mutation operators before the mutation starts.

\newcommand{\oplast}{op_{i-1}}
\newcommand{\opnow}{op_i}
\newcommand{\klast}{k_{i-1}}
\newcommand{\know}{k_i}
\newcommand{\oplist}{\textit{op\_list}}

\begin{algorithm}[tb]

    \KwIn{$op_{i-1}$: the mutation operator used in last iteration}
    \KwIn{\pool: a list of predefined input mutation operators}
    \KwOut{$op_{i}$: the mutation operator for this iteration}
    \tcp{sort the operators in \pool into a list in descending order of the operators' ranking values}
    $\oplist \gets \pool.\text{sort}()$  \label{alg:select:line:sort} \\
    $\klast \gets \oplist.\text{index}(\oplast)$  \label{alg:select:line:getidx} \\
    $p_{accept} \gets 0 $ \\
    \While{random.rand(0, 1) $\geq p_{accept}$ }{
        $\opnow \gets \text{a random operator in \oplist}$ \label{alg:select:line:select} \\
        $\know \gets \text{\oplist.index}(\opnow)$ \\
        $p_{accept}  \gets (1-p)^{\know - \klast} $ \label{alg:select:line:prob} \\
    } \label{alg:select:line:probability_accept}
    \Return $\opnow$ \caption{Mutation Operator Selection}
    \label{alg:select}
\end{algorithm}

To tackle the challenge of selecting effective mutation operators,
\proj models the problem as a Markov Chain~\cite{markovchain} and uses Monte Carlo~\cite{mcmc} to guide the selection.
During the test generation, \proj selects one mutation operator from a pool of operators and applies it to the input.
This process can be modeled as a stochastic process $\left\{op_0, op_1, \cdots, op_t \right\}$, where $op_i$ is the selected operator at $i$-th iteration.
Since the selection of $op_{i+1}$ from all possible states only depends on $op_{i}$~\cite{Athena, lemon, jvmdiff},
this process is a typical Markov Chain.
Given this modeling, \proj further uses Markov Chain Monte Carlo (MCMC)~\cite{mcmc} to guide the selection of mutation operators in order to mimic the selection from the actual probability distribution.

Specifically, \proj adopts Metropolis-Hasting algorithm~\cite{mcmc}, a popular MCMC method to guide the selection of mutation operators from the operator pool.
Throughout all iterations, for operator $op$, \proj associates it with a ranking value:
$$v(op) =
	\frac{N_{i}}{N_{op} + \epsilon}$$
where $N_{op}$ is the number of times that operator $op$ is selected and $N_{i}$ is the number of times that the fitness value of input is increased after applying $op$.
$\epsilon = 1e-7$ is used to avoid division by zero when $N_{op} = 0$.
These numbers are dynamically updated in the generation as shown in \cref{alg:main}, \cref{alg:line:increase}.

The detailed algorithm for the operator selection given the operator at last iteration $op_{i-1}$ in \proj
is shown in \cref{alg:select}.
Based on each operator's ranking value $v$, \proj first sorts the mutation operators
in the descending order of $v$ (\cref{alg:select:line:sort}) and denotes the index of $op_{i-1}$ as $k_{i-1}$ (\cref{alg:select:line:getidx}).
Then \proj selects one mutation operator from the pool (\cref{alg:select:line:select}) and calculates the acceptance probability for $op_{i}$ given $op_{i-1}$ (\cref{alg:select:line:prob}):
$$
P\left(op_{i}|op_{i-1}\right) = (1-p)^{k_{i} - k_{i-1}}
$$
where $p$ is the multiplicative inverse for the number of mutation operators in the pool.
Following the Metropolis-Hasting algorithm, \proj randomly accepts or rejects this mutation operator based on its acceptance probability (\cref{alg:select:line:probability_accept}).
The above process will repeat until one operator is accepted.

 \section{Experiment Design}

In this section, we introduce the design of our evaluation.
In particular, we
aim to answer the following four research questions in our evaluation.

\begin{description}
\item [RQ1] Is \proj effective to find triggering inputs?

\item [RQ2] Is \proj time-efficient and query-efficient to find triggering inputs?

\item [RQ3] What are the effects of the fitness function
and the selection strategy of mutation operator used by \proj in finding triggering inputs?

\item [RQ4] Can \proj facilitate the dissemination of compressed models?

\item [RQ5] Can the triggering input found by \proj be used to repair the deviated behaviors?

\end{description}

We collect \NumOriginalCompressedParis pairs of original models and their compress models to answer the effectiveness and efficiency of \proj in \textbf{RQ1} and \textbf{RQ2}.
For \textbf{RQ3}, we conduct an ablation study to understand
the impacts of our fitness function and mutation operation selection strategy
on effectiveness and efficiency.
In \textbf{RQ4}, we design a case study and discuss one potential application of \proj
to facilitate model dissemination.
For \textbf{RQ5}, we explore the possibility to repair the deviated behaviors of compressed models
using the triggering input found by \proj.

\subsection{Datasets and Seed Inputs}
We use the three datasets: MNIST~\cite{lecun-mnisthandwrittendigit-2010},
CIFAR-10~\cite{cifar10}
and ImageNet~\cite{imagenet_cvpr09}
to evaluate the performance of \proj.
We choose them as they are widely used for image classification tasks, and there are many models trained on them so that we can collect a sufficient number of compressed models for evaluation.
These datasets are also used by many studies in model compression~\cite{7780890,
    deepcompression, compressionsurvery,9384353,10.1145/3489517.3530400,
    DBLP:conf/iclr/CaiGWZH20,9008764,DBLP:conf/iclr/ZhuG18,10.5555/3367471.3367522,DBLP:conf/iclr/LinSBDJ20}.
For each dataset, we randomly select 500 images as seed inputs from their test set for evaluation.
Each seed input in MNIST and CIFAR-10 is pre-processed by normalization based on the mean value $mean$ and standard deviation $std$ of the dataset, \ie, $\frac{x_s- mean}{std}$.
For the inputs in ImageNet, they are pre-processed using the function provided by each model.
To mitigate the impact of randomness, we repeat the experiments five times and each time use a different random seed.

\subsection{Compressed Models}
The compressed models used in our evaluation come from two sources.
First, we use three pairs of the original model and the according quantized model used by \df:
LeNet-1 and LeNet-5 for MNIST, and ResNet-20 for CIFAR-10.
They are compressed by the authors of \df using TensorFlow Lite~\cite{tflite} with 8-bit quantization.
The upper half of \cref{tab:model2} shows their top-1 accuracy.

Second, to comprehensively evaluate the performance of \proj on other kinds of compressed techniques, we also prepare 15 pairs of models.
Specifically, six of them are for MNIST and nine of them are for CIFAR-10.
These compressed models are prepared by three kinds of techniques, namely, quantization, pruning, and knowledge distillation, using Distiller, an open-source model compression toolkit built by the Intel AI Lab~\cite{nzmora2019distiller}.
The remaining three models for ImageNet and their quantized models are collected from PyTorch~\cite{pytorchmodels}.
These three models are chosen since their accuracy is highest among all compressed models in PyTorch Models.
The lower half of \cref{tab:model2} shows their top-1 accuracy.

\begin{table}[t]
	\centering
	{
		\caption{The Top-1 accuracy of the original models and compressed models used in the evaluation.
			The first three models are from \df and the other models are prepared by this study.
		}
		\label{tab:model2}

	\begin{tabular}{@{}ccccc@{}}
				\toprule
                \textbf{Dataset}           & \textbf{Original Model} & \textbf{Accuracy}(\%) & \textbf{Compression Method} & \textbf{Accuracy}(\%) \\ \midrule \midrule
				\multirow{2}{*}{MNIST}   & LeNet-1                 & 97.88               & Quantization-8-bit & 97.88            \\ \cmidrule(l){2-5}
				& LeNet-5                 & 98.81               & Quantization-8-bit & 98.81            \\ \midrule
				CIFAR-10                 & ResNet-20                 & 91.20               & Quantization-8-bit & 91.20            \\ \midrule \midrule
				\multirow{6}{*}{MNIST}    & \multirow{2}{*}{CNN}                                              & \multirow{2}{*}{99.11}                                           & Pruning             & 99.23                                                            \\
				&                                                                   &                                                                  & Quantization             & 99.13                                                            \\ \cmidrule(l){2-5}
				& \multirow{2}{*}{LeNet-4}                                          & \multirow{2}{*}{99.21}                                           & Pruning             & 99.13                                                            \\
				&                                                                   &                                                                  & Quantization             & 99.21                                                            \\ \cmidrule(l){2-5}
				& \multirow{2}{*}{LeNet-5}                                          & \multirow{2}{*}{99.13}                                           & Pruning             & 98.99                                                            \\
				&                                                                   &                                                                  & Quantization             & 99.15                                                            \\ \midrule
				\multirow{9}{*}{CIFAR-10} & \multirow{3}{*}{PlainNet-20}                                      & \multirow{3}{*}{87.33}                                           & Knowledge Distillation             & 75.89                                                            \\
				&                                                                   &                                                                  & Pruning             & 85.98                                                            \\
				&                                                                   &                                                                  & Quantization             & 87.12                                                            \\ \cmidrule(l){2-5}
				& \multirow{3}{*}{ResNet-20}                                        & \multirow{3}{*}{89.42}                                           & Knowledge Distillation            & 74.60                                                            \\
				&                                                                   &                                                                  & Pruning             & 89.88                                                            \\
				&                                                                   &                                                                  & Quantization             & 88.89                                                            \\ \cmidrule(l){2-5}
				& \multirow{3}{*}{VGG-16}                                           & \multirow{3}{*}{87.48}                                           & Knowledge Distillation            & 87.59                                                            \\
				&                                                                   &                                                                  & Pruning             & 88.44                                                            \\
				&                                                                   &                                                                  & Quantization             & 87.06                                                            \\ \midrule
                \multirow{3}{*}{ImageNet} &  \multirow{1}{*}{Inception}  & 93.45                  &      Quantization      & 93.35                  \\
                &  \multirow{1}{*}{ResNet-50}  & 95.43                  &      Quantization      & 94.98                  \\
                & \multirow{1}{*}{ResNeXt-101} & 96.45                  &      Quantization      & 96.33                   \\ \bottomrule

			\end{tabular}}
\end{table}

\subsection{Evaluation Metrics}

For effectiveness, we measure the success rate to find a triggering input for selected seed inputs.
In terms of efficiency, we measure the average time and model queries it takes to find a triggering input for each seed input.
All of them are commonly used by related studies~\cite{diffchaser,simpleattack,deepxplore,tensorfuzz}.
Their details are explained as follows.

\myparagraph{Success Rate}
It measures the ratio of the seed inputs based on which a triggering input is successfully found over the total number of seed inputs.
The higher the success rate, the more effective the underlying methodology.
Specifically, $$\text{Success Rate} = \frac{\sum_{i=1}^{N} {s_{x_i}} }{N}$$
where $s_{x_i}$ is an indicator: it is equal to 1 if a triggering input based on seed input $x_i$ is found.
Otherwise, $s_{x_i}$ is 0.
$N$ is the total number of seed inputs, \ie, 500 in our experiments.

\myparagraph{Average Time}
It is the average time to find a triggering input for each seed input.
Mathematically, $$\text{Average Time} = \frac{\sum_{i=1}^{N} {t_{x_i}}}{N}$$
where $t_{x_i}$ is the time spent to find a triggering input given the seed input $x_i$.
The shorter the time, the more efficient the input generation.
We measure the average time spent to find all triggering inputs provided the seed inputs.

\myparagraph{Average Query}
It measures the average number of model queries issued by \proj to find a triggering input for each seed input.
Formally, this metric is defined as: $$\text{Average Queries} = \frac{\sum_{i=1}^{N} {q_{x_i}}}{N}$$
where $q_{x_i}$ is the number of queries to find a triggering input given the seed input $x_i$.
A model query means that one input is fed into both the original DNN model and the compressed one.
Since the computation of the DNN models is expensive, it is preferred to issue as few queries as possible.
The fewer the average queries, the more efficient the test generation.

\subsection{Experiments Setting}
\label{sec:eval_design}
\myparagraph{Baseline and its Parameters}
We use the \df~\cite{diffchaser} as the baseline, since it is the state-of-the-art black-box approach to our best knowledge.
Specifically, we use the source code and its default settings provided by the corresponding authors.
For the timeout to find triggering inputs for each seed input, we use the same setting as \df, \ie 180s.
The experiment platform is a CentOS server with a CPU 2xE5-2683V4 2.1GHz and a GPU 2080Ti.

\myparagraph{Mutation Operators}
For a fairness evaluation, we used the same image mutation operators from the baseline \df, as shown in \cref{tab:mutop}.
These mutation operators are proposed by prior work~\cite{deepxplore,diffchaser,deeptest, deepgauge,deephunter}
to simulate the scenario that DNN models are likely to face in the real world.
For example, Gaussian Noise is considered as one of the most frequently
occurring noises in image processing~\cite{BONCELET2009143}.
After applying each mutation operator to a given image,
we clip the values of pixels to $[0, 255]$ so that the resulted images are still valid images.
Please note that these mutation operators may have certain randomness.
For example, the size of the average filter used by \textit{Average Blur Image} is
randomly selected from  1 to 5.

\begin{table}[ht]
	\centering
	\caption{Mutation operators used in \proj and \df}
	\label{tab:mutop}
	\resizebox{\linewidth}{!}{\begin{tabular}{@{}lll@{}}
			\toprule
			Category & Mutation Operator &  Description\\ \midrule
			\multirow{3}{*}{Adding Noise} & Random Pixel Change &  Randomly change the values of pixels to arbitrary values in $[0, 255]$\\
			& Gaussian Noise & Generate a random Gaussian-distributed noise~\cite{gonzalez2008digital} and add it into the image. \\
			& Multiplicative Noise  & Generate a random Multiplicative noise~\cite{gonzalez2008digital} and add it into the image  \\ \midrule
			\multirow{3}{*}{Blurring Image} &  Average Blur Image&  Blur the image using a random average filter. \\
			&  Gaussian Blur Image &Blur the image using a random Gaussian filter.  \\
			& Median Blur Image  & Blur the image using a random median filter. \\ \bottomrule
		\end{tabular}
	}
\end{table}  \section{Evaluation Results and Analysis}
\label{sec:eval}

\subsection{RQ1: Effectiveness}

\subsubsection{Triggering Inputs found by \proj}
\begin{figure}[h]
	\centering
	\begin{subfigure}[b]{.20\linewidth}
		\includegraphics[width=\linewidth]{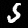}
	\end{subfigure}
	\hfil
	\begin{subfigure}[b]{.20\linewidth}
		\includegraphics[width=\linewidth]{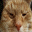}
	\end{subfigure}
	\hfil
	\begin{subfigure}[b]{.20\linewidth}
		\includegraphics[height=\linewidth]{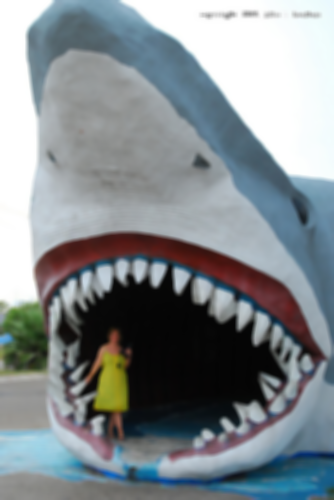}
	\end{subfigure}

	\caption{Triggering Inputs Found by \proj.}
	\label{fig:example}
\end{figure}

\cref{fig:example} shows three examples of the triggering inputs found by \proj in MNIST, CIFAR-10, and ImageNet respectively.
The original models correctly classify the two inputs as ``5'', ``cat'',
``great white shark'' respectively.
However, the inputs are misclassified as ``6'', ``deer'' and ``marimba'' (a musical instrument) by the associated compressed models, respectively.

\subsubsection{Success Rate}
The two \textbf{Average Success Rate} columns in \cref{tab:results1} show the success rate of \proj and \df, respectively.
\proj achieves 100\% success rate for all pairs of models on three datasets.
As for \df,
its success rate on MNIST and CIFAR-10 datasets, ranges from 74.12\% to 99.92\%,
with an average of 96.39\%.
Such results indicate that \df fails to find the triggering input for certain seed inputs of all the pairs.
Specifically, the success rate of \df is lower than 90\% for three Quantization Model in the CIFAR-10 dataset, while \proj constantly achieves 100\% success rate in all models.
For the models that are trained on ImageNet,
the success rates of \df range from 12.01\% to 21.12\%.
This result demonstrates that \proj outperforms \df in terms of effectiveness.
The reason is that \df, especially its \textit{k-Uncertainty} fitness function, does not properly measure the differences between two models, resulting in failures to find triggering input for certain cases.
In contrast, the fitness function of \proj not only measures the differences
between the prediction outputs of the original and compressed models,
but also measures whether the input triggers previously unobserved states of two models.
By combining this fitness function with our advanced selection strategy of mutation operators, our approach always achieves 100\% success rates in our experiments.
\begin{table*}[]
	\centering
	{\caption{Comparison of effectiveness and time-/query-efficiency between \proj and \df.
			The results are averaged across five runs using different random seeds.
		}
		\label{tab:results1}
\resizebox{\textwidth}{!}{\begin{tabular}{@{}ccc|rrr|rrr@{}}
\toprule
\multirow{4}{*}{Dataset}     & \multirow{4}{*}{Model}        & \multirow{4}{*}{Compression}                & \multicolumn{3}{c|}{\proj}                                                                                                                                       & \multicolumn{3}{c}{\df}                                                                                                                                  \\ \cmidrule(lr){4-6}\cmidrule(lr){7-9}
&                               &                                             & \multicolumn{1}{c}{\textbf{Average}} & \multicolumn{1}{c}{\textbf{Average}} & \multicolumn{1}{c|}{\multirow{2}{*}{\textbf{Average}}}  & \multicolumn{1}{c}{\textbf{Average}} & \multicolumn{1}{c}{\textbf{Average}}  & \multicolumn{1}{c}{\multirow{2}{*}{\textbf{Average}}} \\
&                               &                                             & \multicolumn{1}{c}{\textbf{Success}} & \multicolumn{1}{c}{\textbf{Time}} & \multicolumn{1}{c|}{\multirow{2}{*}{\textbf{Query}}}  & \multicolumn{1}{c}{\textbf{Success}} & \multicolumn{1}{c}{\textbf{Time}}  & \multicolumn{1}{c}{\multirow{2}{*}{\textbf{Query}}} \\
&                               &                                             & \multicolumn{1}{c}{\textbf{Rate}} & \multicolumn{1}{c}{\textbf{(sec)}} &  & \multicolumn{1}{c}{\textbf{Rate}} & \multicolumn{1}{c}{\textbf{(sec)}}  &  \\ \midrule
MNIST                     & LeNet-1                       & \multicolumn{1}{c|}{Quantization-8-bit}     & 100\%                                  & 0.513                             & \multicolumn{1}{r|}{83.97}  & {99.40\%}                                & 10.654                                                & 5812.47                                             \\ \cmidrule(l){2-9}
& LeNet-5                       & \multicolumn{1}{c|}{Quantization-8-bit}     & 100\%                                  & 0.706                             & \multicolumn{1}{r|}{117.02} & {99.68\%}                                & 12.598                                                & 6040.53                                             \\ \midrule
CIFAR-10                  & \multicolumn{1}{l}{ResNet-20} & \multicolumn{1}{c|}{Quantization-8-bit}     & 100\%                                  & 0.509                             & \multicolumn{1}{r|}{30.43}  & {99.76\%}                                & 33.980                                                & 2323.58                                             \\ \midrule \midrule
\multirow{6}{*}{MNIST}    & \multirow{2}{*}{LeNet-4}      & \multicolumn{1}{c|}{Prune}                  & 100\%                                  & 0.056                             & \multicolumn{1}{r|}{18.34}  & {99.44\%}                                & 16.249                                                & 6172.57                                             \\
&                               & \multicolumn{1}{c|}{Quantization}           & 100\%                                  & 0.187                             & \multicolumn{1}{r|}{27.83}  & {98.08\%}                                & 76.254                                                & 6506.53                                             \\ \cmidrule(l){2-9}
& \multirow{2}{*}{LeNet-5}                       & \multicolumn{1}{c|}{Prune}                  & 100\%                                  & 0.071                             & \multicolumn{1}{r|}{22.03}  & {98.56\%}                                & 17.446                                                & 6276.38                                             \\
&                               & \multicolumn{1}{c|}{Quantization}           & 100\%                                  & 0.225                             & \multicolumn{1}{r|}{28.08}  & {98.48\%}                                & 45.618                                                & 6662.88                                             \\ \cmidrule(l){2-9}
& \multirow{2}{*}{CNN}          & \multicolumn{1}{c|}{Prune}                  & 100\%                                  & 0.068                             & \multicolumn{1}{r|}{22.51}  & {99.60\%}                                & 16.381                                                & 6053.82                                             \\
&                               & \multicolumn{1}{c|}{Quantization}           & 100\%                                  & 0.173                             & \multicolumn{1}{r|}{25.34}  & {99.52\%}                                & 38.039                                                & 6450.96                                             \\ \midrule
\multirow{9}{*}{CIFAR-10} & \multirow{3}{*}{PlainNet-20}  & \multicolumn{1}{c|}{Prune}                  & 100\%                                  & 0.051                             & \multicolumn{1}{r|}{4.31}   & {99.80\%}                                & 18.222                                                & 1896.59                                             \\
&                               & \multicolumn{1}{c|}{Quantization}           & 100\%                                  & 0.470                             & \multicolumn{1}{r|}{9.13}   & {89.52\%}                                & 75.191                                                & 1696.16                                             \\
&                               & \multicolumn{1}{c|}{Knowledge Distillation} & 100\%                                  & 0.029                             & \multicolumn{1}{r|}{3.97}   & {99.72\%}                                & 12.324                                                & 1961.09                                             \\ \cmidrule(l){2-9}
& \multirow{3}{*}{ResNet-20}    & \multicolumn{1}{c|}{Prune}                  & 100\%                                  & 0.063                             & \multicolumn{1}{r|}{4.70}   & {99.88\%}                                & 23.298                                                & 2145.77                                             \\
&                               & \multicolumn{1}{c|}{Quantization}           & 100\%                                  & 0.685                             & \multicolumn{1}{r|}{10.16}  & {74.12\%}                                & 83.971                                               & 1511.06                                             \\
&                               & \multicolumn{1}{c|}{Knowledge Distillation} & 100\%                                  & 0.032                             & \multicolumn{1}{r|}{3.91}   & {99.92\%}                                & 14.117                                                & 2097.28                                             \\ \cmidrule(l){2-9}
& \multirow{3}{*}{VGG-16}       & \multicolumn{1}{c|}{Prune}                  & 100\%                                  & 0.041                             & \multicolumn{1}{r|}{5.84}   & {99.60\%}                                & 15.619                                                & 2453.01                                             \\
&                               & \multicolumn{1}{c|}{Quantization}           & 100\%                                 & 1.183                             & \multicolumn{1}{r|}{26.16}  & 80.08\%                                & 85.709                                               & 2129.37                                             \\
&                               & Knowledge Distillation                      & 100\%                                                       & 0.036                             & 5.78                        & 99.80\%                                                     & 16.058                                                & 2761.12                                             \\ \midrule
\multirow{3}{*}{ImageNet} &  \multirow{1}{*}{Inception}  &     Quantization               &                              100\% &                              1.266 &                                               21.44 & 20.20\%& 163.847  & 1808.87 \\
&  \multirow{1}{*}{ResNet-50}  &      Quantization                &                                100\% &                              0.819 &                                               19.24 &  21.12\%&158.393 & 1702.10 \\
& \multirow{1}{*}{ResNeXt-101}  &      Quantization                     &                              100\% &                              3.693&                                               34.49  & 12.01\%& 163.936 &2030.41 \\ \bottomrule

\bottomrule
\end{tabular}}
}

\end{table*}

To further investigate the effectiveness of \proj, we feed all the non-triggering inputs in the entire test set as seed inputs into \proj on the \NumOriginalCompressedParis pairs of models.
We found that \proj can consistently achieve a 100\% success rate for all \NumOriginalCompressedParis pairs.
The result on ImageNet models is in shown in \cref{tab:imagenet}
and it shows that \proj is effective to find the triggering inputs
for these large models trained on complex dataset and the success rates are 100\% in five runs.
Due to the low efficiency of \df as shown in the next section, we are not able to conduct the same experiments using \df.
\begin{table*}[h]
\centering
\caption{Effectiveness of \proj on on ImageNet models using entire test set as seed inputs.
The inputs in the ImageNet test set that can trigger deviated behaviors are excluded from experiments.
The results are averaged across five runs.}
\label{tab:imagenet}
\resizebox{\linewidth}{!}{
\begin{tabular}{@{}ccccc|rrr@{}}
	\toprule
	\multirow{3}{*}{Dataset}  &    \multirow{3}{*}{Model}    & \multirow{3}{*}{Accuracy} & \multirow{3}{*}{Compression} & \multirow{3}{*}{Accuracy} & \multicolumn{1}{c}{\textbf{Average}} & \multicolumn{1}{c}{\textbf{Average}} & \multicolumn{1}{c}{\multirow{2}{*}{\textbf{Average}}} \\
	                          &                              &                           &                              &                           & \multicolumn{1}{c}{\textbf{Success}} &    \multicolumn{1}{c}{\textbf{Time}} &   \multicolumn{1}{c}{\multirow{2}{*}{\textbf{Query}}} \\
	                          &                              &                           &                              &                           &    \multicolumn{1}{c}{\textbf{Rate}} &   \multicolumn{1}{c}{\textbf{(sec)}} &                                                       \\ \midrule
	\multirow{3}{*}{ImageNet} &  \multirow{1}{*}{Inception}  & 93.45\%                   &      Quantization      & 93.35\%                  &                              100\% &                              1.22 &                                               18.54 \\
	                          &  \multirow{1}{*}{ResNet-50}  & 95.43\%                  &      Quantization     & 94.98\%                  &                                100\% &                              0.80 &                                               15.19 \\
	                          & \multirow{1}{*}{ResNeXt-101} & 96.45\%                  &      Quantization     & 96.33\%                   &                              100\% &                              4.33 &                                               28.22 \\ \bottomrule
	                                                     \multicolumn{5}{c}{\textbf{Average}}                                                       &                              100\% &                              2.12 &                                               20.65 \\ \bottomrule
\end{tabular}
}
\end{table*}

\answer{Answer to RQ1}{
		\proj is effective in finding triggering inputs for compressed models.
		Specifically, it constantly achieves 100\% success rate in all \NumOriginalCompressedParis pairs of models.
}

\begin{figure*}[t]
    \centering
    \begin{subfigure}[b]{.475\linewidth}
        \includegraphics[width=\linewidth, page=1]{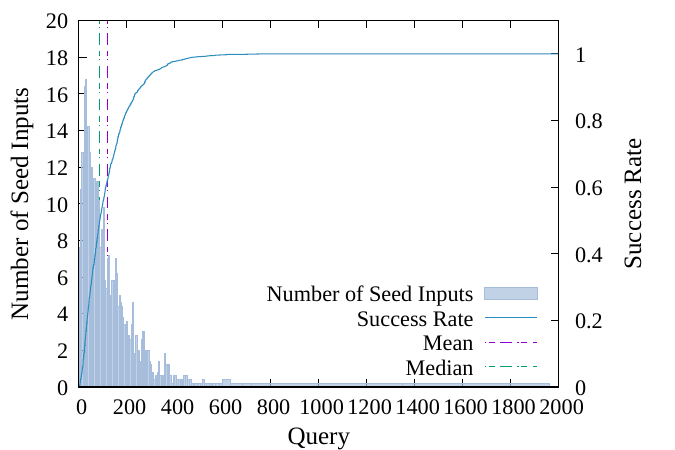}
        \caption{\proj, LeNet-5, Quantization-8-bit}
    \end{subfigure}
    \hfil
    \begin{subfigure}[b]{.475\linewidth}
        \includegraphics[width=\linewidth, page=1]{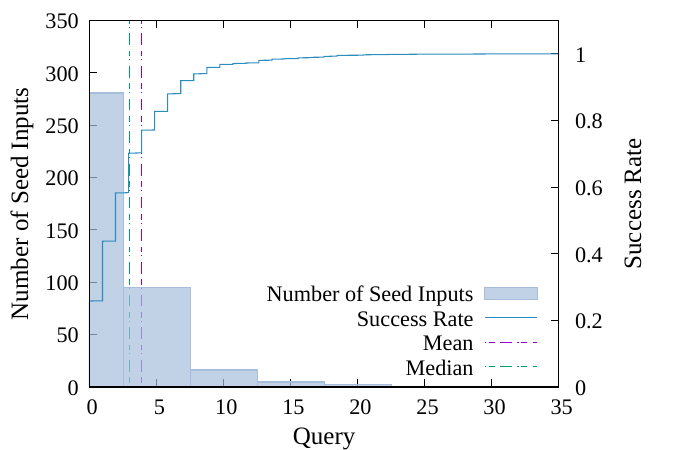}
        \caption{\proj, ResNet-20, Knowledge Distillation}
    \end{subfigure}

    \begin{subfigure}[b]{.475\linewidth}
        \includegraphics[width=\linewidth, page=2]{lenet5-quan.pdf}
        \caption{\df, LeNet5 Quantization-8-bit}
    \end{subfigure}
    \hfil
    \begin{subfigure}[b]{0.475\linewidth}
        \includegraphics[width=\linewidth, page=2]{resnet20-kd.pdf}
        \caption{\df, ResNet-20, Knowledge Distillation}
    \end{subfigure}
    \caption{Histogram of the number of queries required by \proj and \df to find the triggering input for the given seed input.
        The value is averaged over five repeated experiments.}
    \label{fig:query}
\end{figure*}

\subsection{RQ2: Efficiency}

\subsubsection{Time}\label{efficiency:time}
The two \textbf{Average Time} columns in \cref{tab:results1} show the average time spent by \proj and \df to find triggering inputs for each seed input if successful.
The time needed by \proj to find one triggering input ranges from 0.029s to 3.369s, with the average value 0.518s.
\df takes much longer time than \proj.
Specifically,
\df takes 10.654s$\sim$163.936s to find one triggering input, with the average 52.234s.
On average, \proj is 230.94x (17.84x$\sim$446.06x) as fast as \df in terms of time.

\subsubsection{Query}
The two \textbf{Average Query} columns in \cref{tab:results1} show the average query issued by \proj and \df for all seed inputs if a triggering input can be found.
Generally, \proj only needs less than 30 queries to find a triggering input, with only two exceptions.
On average, \proj requires only 24.99 queries (3.9$\sim$117.0).
\df always needs thousands of queries for each trigger input (averagely 3642.50), much more than \proj.
For example, the smallest number of queries needed by \df is 1,896.59 for PlainNet-20 and its pruned model.
In the same pair of models, \proj only needs 4.31 queries on average.
Overall,
the number of queries required by \proj is 0.699\% (0.186\%$\sim$1.937\%) of the one required by \df.

We further visualize the queries of \proj and \df in \cref{fig:query} on two pairs of models: LeNet-5 Quantization-8-bit and ResNet-20 Knowledge Distillation.
They are selected since the ratio of queries needed by \proj over the one needed by \df is the smallest (0.186\%) and largest (1.937\%) in all the \NumOriginalCompressedParis pairs of models.
\cref{fig:query} shows the histogram of the number of queries needed by \proj and \df, respectively, as well as the mean and median.
It can be observed that \proj significantly outperforms \df in terms of queries.
The reason is that \df adopts a genetic algorithm to generate many inputs via crossover and feed them into DNN models in each iteration.
As a result, it requires thousands of queries from the two models to find a triggering input.
In contrast, \proj only needs to generate one mutated input and query once in each iteration.

\answer{Answer to RQ2}{
\proj is efficient to find triggering inputs in terms of both time and queries.
On average, \proj is 230.94x as fast as \df and takes only 0.699\% queries as \df.
}

\subsection{RQ3: Ablation Study}
We further investigate the effects of our fitness function and mutation operator selection strategy.
Specifically, we create the following two variants of \proj.

\begin{enumerate}
    \item \projd: the fitness function in \proj is replaced by a simpler fitness function:
    $H_{f,g}(x)=\mathcal{D}_{f,g}(x)=|p_{f(x)} - p_{g(x)}|$.
    In other words, the fitness function does not trace the model states triggered by inputs.
    \item \projr: the selection strategy for mutation operators in \proj is changed to uniform random selection.
\end{enumerate}

\begin{table*}[]
	\centering
	{\caption{Evaluation results of  \projd and \projr.
			The numbers in parentheses are the ratios of time or queries spent by each variant with respect the one spent by \proj.
			The results are averaged across five runs using different random seeds.
		}
		\label{tab:ablation}
\resizebox{\textwidth}{!}{\begin{tabular}{@{}ccc|rrr|rrr@{}}
				    \toprule
				          \multirow{4}{*}{Dataset}       &    \multirow{4}{*}{Model}     &        \multirow{4}{*}{Compression}         &                                                                \multicolumn{3}{c|}{\projd}                                                                 &                                                          \multicolumn{3}{c}{\projr}                                                          \\
				    \cmidrule(lr){4-6}\cmidrule(lr){7-9} &                               &                                             & \multicolumn{1}{c}{\textbf{Average}} &            \multicolumn{1}{c}{\textbf{Average}} &            \multicolumn{1}{c|}{\multirow{2}{*}{\textbf{Average}}} & \multicolumn{1}{c}{\textbf{Average}} &          \multicolumn{1}{c}{\textbf{Average}} & \multicolumn{1}{c}{\multirow{2}{*}{\textbf{Average}}} \\
				                                         &                               &                                             & \multicolumn{1}{c}{\textbf{Success}} &               \multicolumn{1}{c}{\textbf{Time}} &              \multicolumn{1}{c|}{\multirow{2}{*}{\textbf{Query}}} & \multicolumn{1}{c}{\textbf{Success}} &             \multicolumn{1}{c}{\textbf{Time}} &   \multicolumn{1}{c}{\multirow{2}{*}{\textbf{Query}}} \\
				                                         &                               &                                             &    \multicolumn{1}{c}{\textbf{Rate}} &              \multicolumn{1}{c}{\textbf{(sec)}} &                                                                   &    \multicolumn{1}{c}{\textbf{Rate}} &            \multicolumn{1}{c}{\textbf{(sec)}} &                                                       \\ \midrule
				                   MNIST                 &            LeNet-1            &   \multicolumn{1}{c|}{Quantization-8bit}    &                              41.44\% &  39.988                                 (77.93) &   \multicolumn{1}{r|}{8140.54                            (96.94)} &                                100\% & 0.537                                  (1.05) &             89.52                              (1.07) \\ \midrule
				            \multicolumn{1}{l}{}         &            LeNet-5            &   \multicolumn{1}{c|}{Quantization-8bit}    &                              39.44\% &  38.244                                 (54.15) &   \multicolumn{1}{r|}{7422.38                            (63.43)} &                                100\% & 0.752                                  (1.07) &             125.87                             (1.08) \\ \midrule
				        \multicolumn{1}{l}{CIFAR-10}     & \multicolumn{1}{l}{ResNet-20} &   \multicolumn{1}{c|}{Quantization-8bit}    &                                100\% &   3.454                                  (6.78) &    \multicolumn{1}{r|}{203.91                             (6.70)} &                                100\% & 0.555                                  (1.09) &             30.59                              (1.01) \\ \midrule\midrule
				           \multirow{6}{*}{MNIST}        &   \multirow{2}{*}{LeNet-4}    &         \multicolumn{1}{c|}{Prune}          &                              92.12\% & 8.242                                  (148.23) & \multicolumn{1}{r|}{2970.80                             (161.98)} &                                100\% & 0.062                                  (1.12) &             20.79                              (1.13) \\
				                                         &                               &      \multicolumn{1}{c|}{Quantization}      &                              60.76\% & 31.162                                 (166.55) &  \multicolumn{1}{r|}{4861.92                            (174.69)} &                                100\% & 0.200                                  (1.07) &             30.06                              (1.08) \\
				             \cmidrule(l){2-9}           &            LeNet-5            &         \multicolumn{1}{c|}{Prune}          &                              93.52\% & 7.428                                  (104.62) &  \multicolumn{1}{r|}{2473.71                            (112.27)} &                                100\% & 0.080                                  (1.13) &             24.77                              (1.12) \\
				                                         &                               &      \multicolumn{1}{c|}{Quantization}      &                              59.92\% & 26.699                                 (118.72) &  \multicolumn{1}{r|}{3548.56                            (126.38)} &                                100\% & 0.237                                  (1.05) &             29.99                              (1.07) \\
				             \cmidrule(l){2-9}           &     \multirow{2}{*}{CNN}      &         \multicolumn{1}{c|}{Prune}          &                              81.04\% & 11.396                                 (168.58) &  \multicolumn{1}{r|}{4348.22                            (193.20)} &                                100\% & 0.075                                  (1.11) &             25.14                              (1.12) \\
				                                         &                               &      \multicolumn{1}{c|}{Quantization}      &                              63.68\% & 26.639                                 (154.16) &  \multicolumn{1}{r|}{4243.05                            (167.46)} &                                100\% & 0.182                                  (1.05) &             26.39                              (1.04) \\ \midrule
				         \multirow{9}{*}{CIFAR-10}       & \multirow{3}{*}{PlainNet-20}  &         \multicolumn{1}{c|}{Prune}          &                                100\% &  0.037                                   (1.28) &    \multicolumn{1}{r|}{5.16                               (1.30)} &                                100\% & 0.031                                  (1.05) &             4.05                               (1.02) \\
				                                         &                               &      \multicolumn{1}{c|}{Quantization}      &                                100\% &   0.058                                  (1.14) &    \multicolumn{1}{r|}{5.11                               (1.18)} &                                100\% & 0.051                                  (1.00) &             4.32                               (1.00) \\
				                                         &                               & \multicolumn{1}{c|}{Knowledge Distillation} &                                100\% &   0.809                                  (1.72) &    \multicolumn{1}{r|}{15.85                              (1.74)} &                                100\% & 0.477                                  (1.02) &             9.30                               (1.02) \\
				             \cmidrule(l){2-9}           &  \multirow{3}{*}{ResNet-20}   &         \multicolumn{1}{c|}{Prune}          &                                100\% &   0.039                                  (1.22) &    \multicolumn{1}{r|}{4.95                               (1.27)} &                                100\% & 0.031                                  (0.98) &             3.80                               (0.97) \\
				                                         &                               &      \multicolumn{1}{c|}{Quantization}      &                                100\% &   0.080                                  (1.26) &    \multicolumn{1}{r|}{6.07                               (1.29)} &                                100\% & 0.066                                  (1.03) &             4.89                               (1.04) \\
				                                         &                               & \multicolumn{1}{c|}{Knowledge Distillation} &                                100\% &    1.329                                 (1.94) &    \multicolumn{1}{r|}{19.39                              (1.91)} &                                100\% & 0.690                                  (1.01) &             10.07                              (0.99) \\
				             \cmidrule(l){2-9}           &    \multirow{3}{*}{VGG-16}    &         \multicolumn{1}{c|}{Prune}          &                                100\% &   0.047                                  (1.28) &    \multicolumn{1}{r|}{7.86                               (1.36)} &                                100\% & 0.035                                  (0.95) &             5.49                               (0.95) \\
				                                         &                               &      \multicolumn{1}{c|}{Quantization}      &                                100\% &   0.050                                  (1.20) &    \multicolumn{1}{r|}{7.48                               (1.28)} &                                100\% & 0.042                                  (1.01) &             5.90                               (1.01) \\
				                                         &                               & \multicolumn{1}{c|}{Knowledge Distillation} &                              99.04\% &   4.986                                  (4.22) &    \multicolumn{1}{r|}{117.08                             (4.48)} &                                100\% & 1.133                                  (0.96) &             24.75                              (0.95) \\ \midrule
				         \multirow{3}{*}{ImageNet}       &  \multirow{1}{*}{Inception}   &             Quantization             &                               79.4\% &                                            16.064 (12.69) &                                                             251.55(11.73) &                          100\%                 &                                   1.459	(1.15)	                                        &                                                       25.01	(1.17)                                                                      \\
				                                         &  \multirow{1}{*}{ResNet-50}   &             Quantization              &                               87.4\% &                                            12.789 (15.63) &                                                             253.57(13.18) &                     100\%                 &    1.418	(1.73)	& 20.34	(1.06)    \\
				                                         & \multirow{1}{*}{ResNeXt-101}  &             Quantization              &                               48.2\% &                                            29.019	(7.86) &	277.17(8.04)&                                 100\%                 &                      4.240	(1.15)	& 40.64	(1.18)                                                                                  \\ \bottomrule\bottomrule
				                                      \multicolumn{3}{c}{Average Ratio \wrt \proj }                                    &                                      &                                          50.06 &                                                             54.85 &                                      &                                          1.09 &                                                  1.05 \\ \bottomrule
				\end{tabular}}
		}

	\end{table*}

For each variant, we measure its success rate, computation time, and the number of queries needed using the seed inputs of the preceding experiments.
\cref{tab:ablation} shows the results.
The numbers in parentheses are the ratios of time or queries spent by each variant with respect the one(s) spent by \proj.

\subsubsection{Fitness Function}
The column \textbf{\projd} in \cref{tab:ablation} shows the evaluation results of \projd.
Although \projd still achieves 100\% success rate in half of the \NumOriginalCompressedParis model pairs,
the success rates of \projd for the remaining 21 pairs
are clearly lower than those of \proj, ranging from 39.44\% to 99.04\%.
The average success rate of \projd over all \NumOriginalCompressedParis pairs of models is only 83.14\%.
In terms of the computation time and the number of queries, \projd is much less efficient than \proj.
Specifically, the time spent by \projd is 1.140x$\sim$168.58x of that spent by \proj, with an average value 50.06x.
As for the number of queries needed, the ratios range from 1.18x to 193.20x, and the average ratio is 54.85x.
This result indicates the importance of encouraging the mutated inputs to explore more model states as formulated by our fitness function.

\subsubsection{Selection Strategy of Mutation Operator}
The column \textbf{\projr} in \cref{tab:ablation} shows the evaluation results of \projr.
Same as \proj, \projr achieves 100\% success rate.
In terms of efficiency, the average time spent by \projr is 1.09x (0.95x$\sim$1.73x) of that spent by \proj.
The ratio of queries required by \projr over those by \proj is also 1.05x, ranging from 0.95x to 1.18x.
In 17 out of \NumOriginalCompressedParis pairs,
the time and queries required by \proj are 91.33\% of that required by \projr.
For the remaining 4 pairs, \projr is marginally (3.5\%) more efficient than \proj
in terms of time and the number of queries.
A possible reason is that
with our fitness function, a triggering input for these four pairs can be found in just a few iterations.
In such cases, the selection strategy of \proj has not obtained enough samples
to capture the knowledge of each mutation operator
before the triggering input is found.
Therefore, it is possible that \projr, which adopts a random mutation strategy
with our effective fitness function, can find the triggering inputs sooner.

To check whether \proj statistically outperforms \projr in terms of time,
we conduct Wilcoxon significant test~\cite{wilcoxontest}
and the p-value is $3.604\times10^{-4}$.
The p-value indicates that our MH algorithm for mutation operator selection significantly improves the efficiency of finding triggering inputs.

\answer{Answer to RQ3}{
    Our fitness function and selection strategy both contribute to the effectiveness and efficiency of \proj.
}

\subsection{Application of \proj: Facilitating Model Dissemination}
In this case study, we discuss a potential application of \proj to facilitate model dissemination.
Specifically, we are going to show that, to a certain extent, the time and number of queries taken to find triggering inputs
can be leveraged as
an approximation
of to what extent the behavior of compressed models differs from that of the original models in the dissemination.
Since \proj can provide this metric effectively and efficiently,
we argue that \proj is able to provide developers with in-time feedback complementary to the accuracy metric,
to assess compressed models.

\subsubsection{Correlation}
We would like to understand the correlation between \textit{the time and queries} and \textit{to what extent the behavior of compressed models differs from that of the original models in deployment}.
We manually constructed a series of models from the original models LeNet-5, ResNet-20 and ResNet-50
in~\cref{tab:model2} by mutating $x\%$ of weights, where $x$ ranges from 10 to 50, with a step of 10.
In the mutation, we randomly mutated the $x\%$ of the weights by increasing or decreasing their values by 10\%.
Intuitively, the larger $x$ is, the more likely the behavior of the resulted model differs from the one of original model.
These models serve as a benchmark with the ground truth, \ie, to what extent the resulted model differs from the original one, for our study.
Then we applied \proj using the same experiment settings in \cref{sec:eval_design} and measured the time and number of queries.

\begin{figure}[h]
    \centering
    \begin{subfigure}[b]{.32\linewidth}
        \includegraphics[width=\linewidth,page=1]{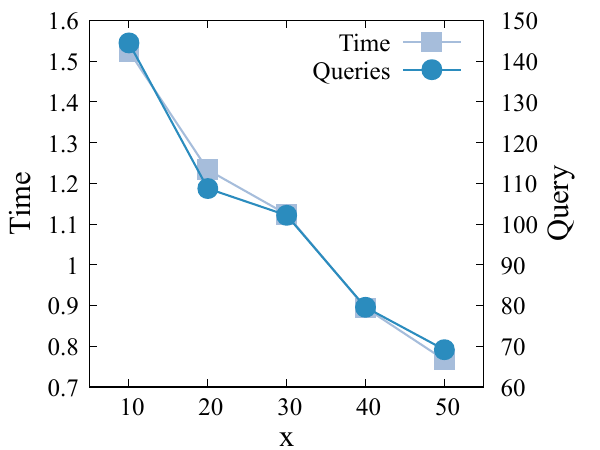}
        \caption{LeNet-5, MNIST}
        \label{fig:corr-lenet}
    \end{subfigure}
    \begin{subfigure}[b]{.32\linewidth}
        \includegraphics[width=\linewidth,page=2]{corr.pdf}
        \caption{ResNet-20, CIFAR-10}
        \label{fig:corr-resnet}
    \end{subfigure}
    \begin{subfigure}[b]{.32\linewidth}
        \includegraphics[width=\linewidth,page=3]{corr.pdf}
        \caption{ResNet-50, ImageNet}
        \label{fig:corr-imagenet}
    \end{subfigure}
    \caption{Correlation between time/query and mutation ratio $x$, using LeNet-5 for MNIST (\cref{fig:corr-lenet}), ResNet-20 for CIFAR-10 (\cref{fig:corr-resnet}), and ResNet-50 for ImageNet (\cref{fig:corr-imagenet}).
    }
    \label{fig:corr}
\end{figure}

\cref{fig:corr-lenet,fig:corr-resnet,fig:corr-imagenet}  show the results for LeNet-5, ResNet-20, and ResNet-50, respectively.
Success rates are not presented since all of them are 100\%.
It is clear that as the portion of the mutated weight $x$\% increases, the time and queries required to find the triggering inputs decrease.
The Pearson Correlation Coefficients~\cite{benesty2009pearson} between $x$ and time/queries also confirm this strong negative correlation, which are -0.989 (time) and -0.972 (queries) for LeNet-5, -0.968 and -0.967 for ResNet-20, and -0.979 and -0.977 for ResNet-50, respectively.
Since the higher $x$ causes the resulted model to be more likely to differ from the original models, we claim that the time and queries can approximate to what extent the behavior of compressed models differs from the one of original models.
Specifically, the less time and fewer queries needed to find triggering inputs, the more likely the compressed model differs from the original model in the dissemination.

\subsubsection{Application}
Now we present an application of \proj in model dissemination.
When compressing a pre-trained model,
developers often need to prepare a compression configuration ~\cite{pruning, compressionsurvery}.
For example, the configuration of model pruning usually specifies which layers in the original model are to be pruned.
A common way is to select the configuration that produces the highest accuracy on test set.
However, as we will demonstrate, only using accuracy is insufficient to distinguish different models, and \proj can provide complementary information to facilitate this process.

\begin{figure}[h]
    \centering
    \includegraphics[width=0.70\linewidth]{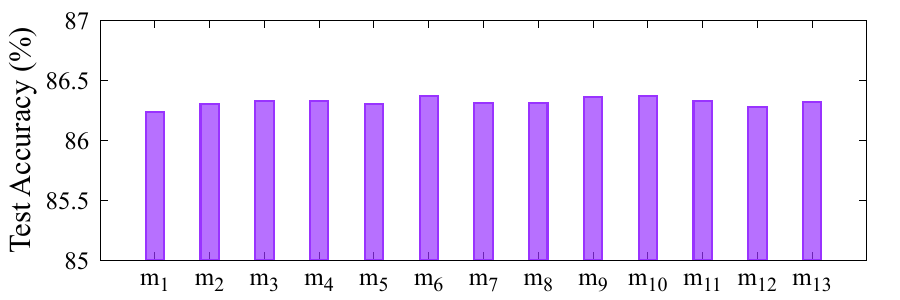}
    \caption{The test accuracy of thirteen compressed models.
    }
    \label{fig:corr-acc}
\end{figure}

We prepared a VGG-16 model by training it from scratch using the CIFAR-10 training dataset.
After the loss and accuracy became saturated, its top-1 accuracy on the CIFAR-10 test set is 86.34\%.
Given this original model, we created a set of compressed models by pruning only one of the
thirteen convolutional layers in the VGG-16 model at one time.
In total, we collected thirteen compressed models using PyTorch
and we referred to them as $m_1$, $m_2$, $\cdots$, $m_{13}$,
where $m_i$ is the compressed model
obtained by pruning the $i$-th convolutional layer of the original model.
\cref{fig:corr-acc} shows the top-1 accuracy of each compressed model.
The accuracy of these models ranges from 86.24\% to 86.37\% and
is almost identical to the accuracy of the original model (86.34\%) with a maximal difference of 0.10\%.
If this developer uses accuracy as the single evaluation metric,
it seems that these models achieve indistinguishable performance,
and thus it makes no difference to select any of them for dissemination.

\begin{figure}[h]
    \centering
    \begin{subfigure}[b]{.70\linewidth}
        \includegraphics[width=\linewidth,page=1]{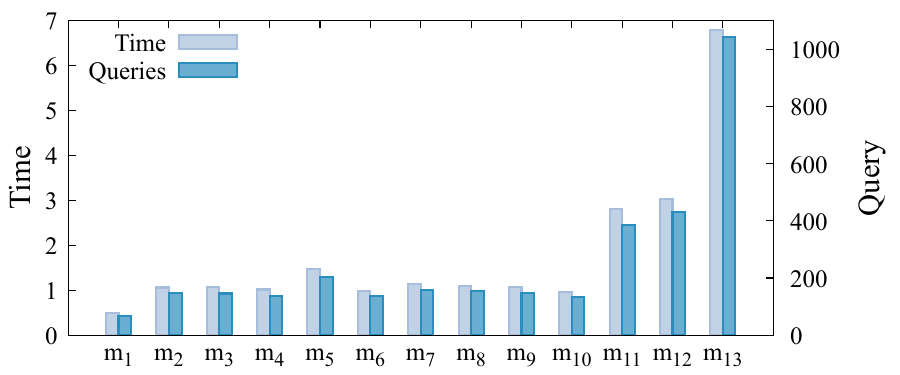}
        \caption{\proj}
        \label{fig:corr-proj}
    \end{subfigure}
    \begin{subfigure}[b]{.70\linewidth}
        \includegraphics[width=\linewidth,page=2]{application.pdf}
        \caption{\df}
        \label{fig:corr-diffchaser}
    \end{subfigure}
    \caption{The results using \proj and \df on thirteen compressed models.
    }
    \label{fig:corr-app}
\end{figure}

In this scenario, \proj can quickly provide
complementary information that is orthogonal to accuracy.
\cref{fig:corr-proj} shows the average time and queries when using \proj to find one bug-triggering input.
Same as the previous settings, we repeated each experiment five times using 500 seed inputs.
Although the accuracy of these models is similar,
the information generated by \proj leads to a different conclusion.
Specifically, it is relatively harder to find a deviated behavior for the compressed model
whose pruned layer is at the bottom of VGG-16, than the models whose pruned layer is at the top.
For example, $m_{13}$ requires much more time and queries than $m_1$.
According to the aforementioned correlation,
if we use the time and number of queries as an approximation
of the likelihood that the compressed model behaves differently from the original model,
it is clear that $m_{13}$ has the least likelihood among all thirteen models.
Taking account of the perspectives from both accuracy and this likelihood information provided by \proj,
the developers should choose the compressed model $m_{13}$ or $m_{12}$ for dissemination,
since they have not only the comparable accuracy, but also the least likelihood to exhibit deviated behaviors.

\cref{fig:corr-diffchaser} shows the results generated by \df.
The average success rate of \df is only 86.3\%, which is 13.7\% lower than \proj.
The time and number of queries required by \df demonstrate the same trend as the one using \proj,
\ie, the models whose pruned layers are at the bottom of the VGG16,
\eg $m_{13}$/$m_{12}$, are less likely to have deviated behaviors than others, \eg $m_1$/$m_2$.
\proj can provide such in-time feedback to developers due to its high effectiveness and efficiency, making it practical to utilize this technique in daily tasks.
In contrast, even though \df may also provide similar information, it takes much a longer time (37.4x on average) and more queries (29.74x) to do so, imposing large computation cost.
For example, given a set of 500 seed inputs and $m_2$, \df requires 6.1 hours and 2,370,800 queries, while \proj only needs 8.9 minutes and 73,320 queries.

\answer{Answer to RQ4}{
    \proj can provide developers with in-time feedback complementary to the accuracy metric,
    to assess compressed models.
}

\subsection{Application of \proj: Repairing the Deviated Behaviors}
\label{subsec:repair}
We further explored the possibility to repair the deviated behaviors
of the compressed models for image classification models
using the triggering inputs found by \proj.
A common approach to improving the performance of DNN models is to retrain the DNN models.
For example, adversarial training can improve the robustness of DNN models~\cite{DBLP:conf/icml/ZhangYJXGJ19, shafahi2019adversarial}.
However, without accessing the internal architectures and status of compressed models,
it is difficult to repair the deviated behaviors directly via retraining.
Therefore, we explored an alternative approach that repairs the deviated behaviors
without the need to retrain the compressed model.
Please note that we are not attempting to repair the triggering inputs in the original test sets,
since the number of triggering inputs in the original test set is ineligible, as shown in~\cref{alg:select}.
It is the duty of compression techniques to reduce the number of triggering inputs in the original test set,
to avoid accuracy degradation due to model compression.

\begin{figure}[h]
    \begin{subfigure}[b]{\linewidth}
        \centering
        \includegraphics[width=\linewidth]{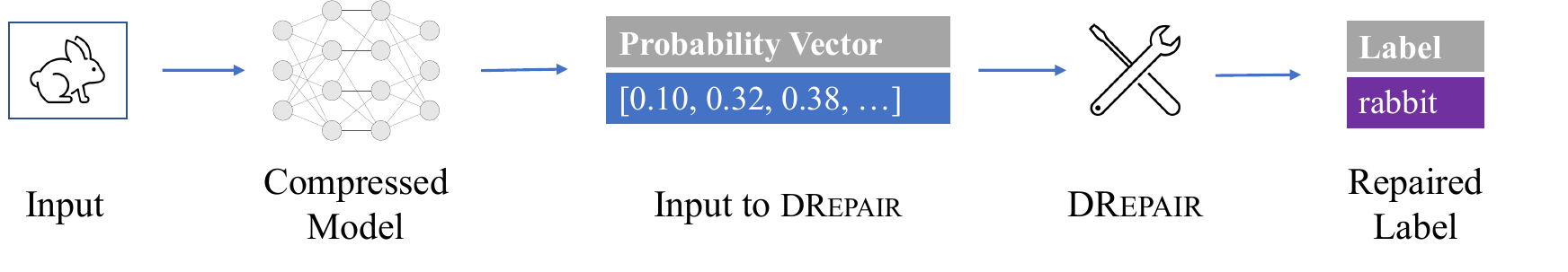}
        \caption{Deployment Stage}
        \label{fig:repair_test}
    \end{subfigure}

    \begin{subfigure}[b]{\linewidth}
        \centering
        \includegraphics[width=\linewidth]{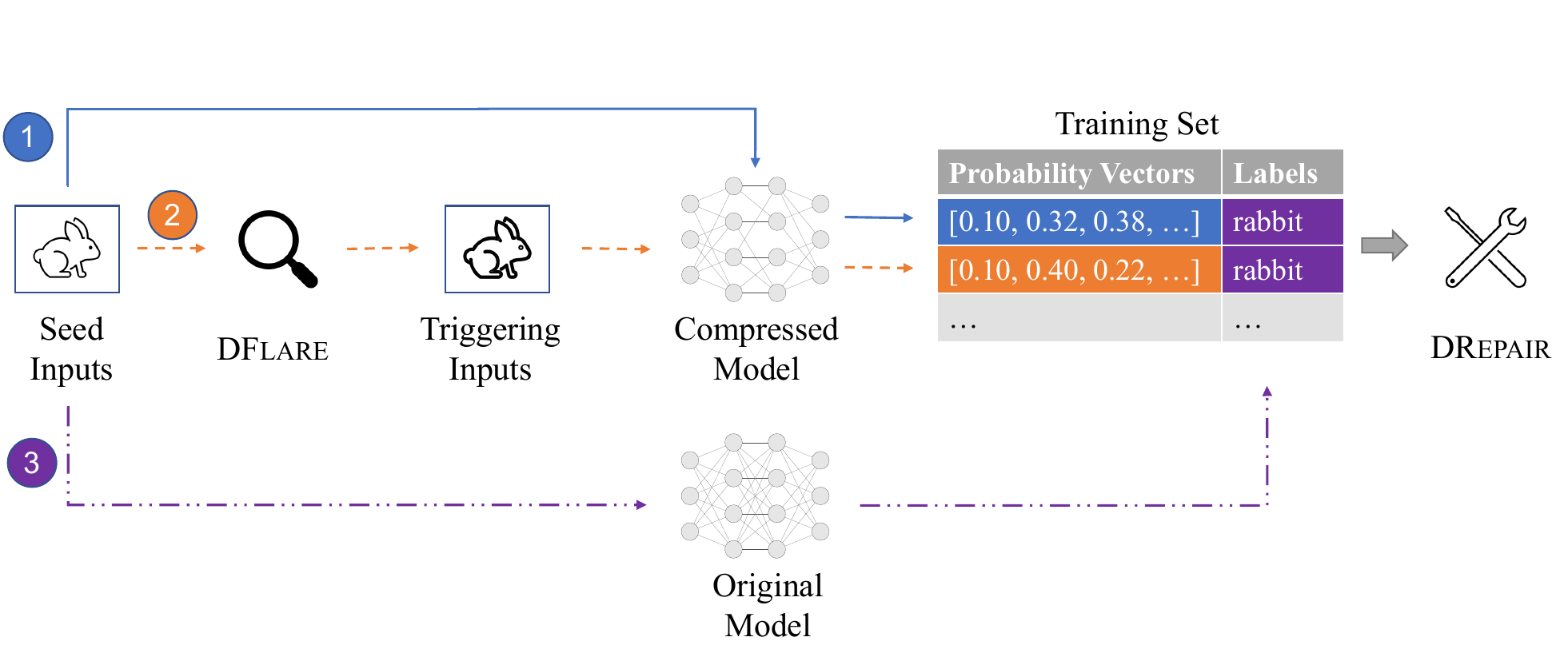}
        \caption{Training Stage}
        \label{fig:repair_train}
    \end{subfigure}
    \caption{The workflow of \repair to repair deviated behaviors of compressed model.}
    \label{fig:repair}
\end{figure}

\subsubsection{Design of \repair}
We proposed a prototype, \repair, to repair the deviated behaviors of the compressed models
for image classifications.
Our intuition is that the substantial amount of triggering inputs found by \proj
contains essential characteristics of such triggering inputs,
and may thus be used to train a separate repair model to fix the deviated behaviors.
\cref{fig:repair_test} illustrates the workflow of \repair.
\repair is a supervised classifier, serving as a post-processing stage of the target compressed model.
Given an input $x$ and the probability vector
$g(x) = [p_1, p_2, p_3, \cdots, p_n]$ outputted by a compressed model $g$,
\repair takes as input the probability vector $g(x)$ and
is expected to output a label $\overline{l_{g(x)}}$ such that
$\overline{l_{g(x)}} = l_{f(x)}$,
where the label $l_{f(x)}$ is outputted by the original model $f$ given the same input $x$.

\cref{fig:repair_train} shows the workflow to train \repair.
After collecting a set of seed inputs,
we first \circled{1} feed each seed input $x_s$  to the compressed model under test $g$ and
collect the probability vector $g(x_s)$.
Then we \circled{2} utilize \proj to find the triggering input $x_t$ given the seed input $x_s$
and obtain its probability vector $g(x_t)$ using the compressed model $g$.
Since \repair is a supervised classifier, each vector in the training set is assigned a target label.
For vector $g(x)$, we \circled{3} use the label outputted by the original model $f$ given input $x$ as the target label,
\ie $l_{f(x)}$.
This is because the objective of \repair is to produce a label that is the same as the label from original model.

\subsubsection{Implementation and Evaluation of \repair}
We implemented \repair using a Single-layer Perceptron (SLP),
\ie, a neural network with a single hidden layer~\cite{slp}.
We chose SLP since it is light-weight in terms of computational resources and
thus is applicable to be deployed along with compressed models in
embedded systems.
We used five-fold cross-validation to evaluate the performance of \repair using the seed inputs and triggering inputs found by \proj in RQ2.
Specifically, for each set of 500 pairs of seed input and triggering input,
we collected their probability vectors and split them into five portions of equal size.
We chose four of them for the training set of \repair, and the remaining one as its test set.
In other words, each training set contains 400 non-triggering inputs and 400 triggering inputs,
and each test set $X$ contains 100 non-triggering inputs and 100 triggering inputs.
In a five-fold cross-validation, we repeated the training and testing five times and ensured a different training and test set is used in each time.
Each five-fold cross-validation was conducted 5 times using different random seeds.

We measure the performance of \repair from the following three perspectives.
In particular, we use $X_t$ to denote the set of triggering input and
use ${X_s}$ to denote non-triggering inputs in the test set $X$.
As we mentioned in the above paragraph,
the sizes of $X_t$, $X_s$ and $X$ are 100, 100 and 200, respectively.

\mymetircs{Repair Count}{Repair Ratio}
We first use \textit{Repair Count}
to measure the number of triggering inputs in $X_t$ that do not trigger deviated behaviors \textit{after} repair,
\ie, $$\text{Repair Count} = \sum_{i=1}^{|X_t|} {o_{x_i}}$$
\noindent where $o_{x_i}$ is an indicator and $o_{x_i} = 1$ only if $\overline{l_{g(x_i)}} = l_{f(x_i)}$,
\ie, $x_i$ does not trigger deviated behavior after repair; otherwise, it is 0.
$|X_t|$ is the number of inputs in $X_t$.

We then measure \textit{Repair Ratio}, \ie, the ratio of Repair Count in $X_t$.
\textit{Repair Ratio} measures the percentage of triggering inputs in $X_t$ that do not trigger deviated behaviors after repair.
The higher the repair ratio is, the more triggering inputs are repaired by \repair.
$$\text{Repair Ratio} =\frac{\text{Repair Count}}{|X_t|} \times 100\% = \frac{\sum_{i=1}^{|X_t|} {o_{x_i}} }{|X_t|} \times 100\%$$

\mymetircs{Inducing Count}{Inducing Ratio}
We use \textit{Inducing Count} to denote the number of non-triggering inputs in
$X_s$ that trigger deviated behaviors \textit{after} repair, \ie
$$\text{Inducing Count} = \sum_{i=1}^{|X_s|} {k_{x_i}}$$
\noindent where $k_{x_i}$ is an indicator and $k_{x_i} = 1$ only if $\overline{l_{g(x_i)}} \neq l_{f(x_i)}$,
\ie, $x_i$ triggers deviated behaviors after repair;
otherwise, it is 0.
$|X_s|$ is the number of inputs in  $X_s$.

Then we use \textit{Inducing Ratio} to measure the ratio of $\text{Inducing Count}$ in $X_s$.
Specifically, \textit{Inducing Ratio} measure the percentage of non-triggering inputs in $X_s$
that trigger deviated behaviors after repair.
The lower the inducing ratio is, the fewer deviated behaviors are induced by \repair.
$$\text{Inducing Ratio} = \frac{\text{Inducing Count}}{|X_s|}= \frac{\sum_{i=1}^{|X_s|} {k_{x_i}} }{|X_s|} \times 100\%$$

\mymetircs{Improvement Count}{Improvement Ratio}
We use \textit{Improvement Count} to measure the difference between
the number of deviated behaviors in $X$ \textit{before} repair by \repair and
the number of deviated behaviors in $X$ \textit{after} repair.
Specifically, the number of deviated behaviors \textit{before} the repair is equal
to the number of triggering inputs in $X$, \ie $|X_t|$.
A deviated behavior \textit{after} repair is triggered by ${x_i}$ if $\overline{l_{g(x_i)}} \neq l_{f(x_i)}$.
Since the indicator $k_{x_i} = 1$ if and only if $\overline{l_{g(x_i)}} \neq l_{f(x_i)}$,
the number of deviated behaviors \textit{after} repair  in $X$ is counted as  $\sum_{i=1}^{|X|} {k_{x_i}}$.
Therefore, the difference between the number of deviated behaviors in $X$ before repair
and after repair is denoted as $$\text{Improvement Count} = |X_t| - \sum_{i=1}^{|X|} {k_{x_i}} $$

We then use \textit{Improvement Ratio} to measure the ratio of $\text{Improvement Count}$ \wrt
to the number of deviated behaviors in $X$ \textit{before} repair.
The higher the improvement ratio is, the more effective \repair is to repair the deviated behaviors of compressed models.

$$\text{Improvement Ratio} = \frac{\text{Improvement Count}}{|X_t|} \times 100\%
= \frac{|X_t| - \sum_{i=1}^{|X|} {k_{x_i}} }{|X_t|} \times 100\%$$

Noticed that the Improvement Ratio can be zero or negative when the number of triggering inputs after repair
is equal to or larger than the number of triggering inputs before repair,
\ie, $\text{Improvement Count} \leq 0$.
In such situations, the repair process fails since
the number of deviated behavior after repair is
more than or equal to the number of deviated behavior before repair.

\begin{table*}[h]
	\centering
	\caption{Evaluation results of \repair.
		The results are averaged across five-fold cross-validation.
	Noticed that since $X_s$ and $X_t$ are equal to 100 in each fold of validation, the $\text{Count} = \text{Ratio} \times 100$.
	}
	\label{tab:repair}
\resizebox{\textwidth}{!}{\begin{tabular}{@{}ccc|rrrrrr@{}}
				\toprule
				\multirow{4}{*}{\textbf{Dataset}} & \multirow{4}{*}{\textbf{Model}} &    \multirow{4}{*}{\textbf{Compression}}    &                                                                 \multicolumn{6}{c}{\textbf{\repair}}                                                                  \\
				\cmidrule(l){4-9}         &                                 &                                             &              \multicolumn{2}{c}{Average}              &              \multicolumn{2}{c}{Average}              &              \multicolumn{2}{c}{Average}              \\
				&                                 &                                             &              \multicolumn{2}{c}{Repair}               &             \multicolumn{2}{c}{Inducing}              &            \multicolumn{2}{c}{Improvement}            \\
				&                                 &                                             & \multicolumn{1}{c}{Count} & \multicolumn{1}{c}{Ratio} & \multicolumn{1}{c}{Count} & \multicolumn{1}{c}{Ratio} & \multicolumn{1}{c}{Count} & \multicolumn{1}{c}{Ratio} \\ \midrule
				MNIST               &             LeNet-1             &   \multicolumn{1}{c|}{Quantization-8-bit}   &                   30.56 &                   30.56\% &                    0.36 &                    0.36\% &                   30.20 &                   30.20\% \\
				\cmidrule(l){2-9}         &             LeNet-5             &   \multicolumn{1}{c|}{Quantization-8-bit}   &                   24.28 &                   24.28\% &                    0.12 &                    0.12\% &                   24.16 &                   24.16\% \\ \midrule
				CIFAR-10              &  \multicolumn{1}{l}{ResNet-20}  &   \multicolumn{1}{c|}{Quantization-8-bit}   &                   15.04 &                   15.04\% &                    0.52 &                    0.52\% &                   14.52 &                   14.52\% \\ \midrule\midrule
				\multirow{6}{*}{MNIST}       &    \multirow{2}{*}{LeNet-4}     &         \multicolumn{1}{c|}{Prune}          &                   48.48 &                   48.48\% &                    0.00 &                    0.00\% &                   48.48 &                   48.48\% \\
				&                                 &      \multicolumn{1}{c|}{Quantization}      &                   35.68 &                   35.68\% &                    0.00 &                    0.00\% &                   35.68 &                   35.68\% \\
				\cmidrule(l){2-9}         &    \multirow{2}{*}{LeNet-5}     &         \multicolumn{1}{c|}{Prune}          &                   43.20 &                   43.20\% &                    0.08 &                    0.08\% &                   43.12 &                   43.12\% \\
				&                                 &      \multicolumn{1}{c|}{Quantization}      &                   38.68 &                   38.68\% &                    0.00 &                    0.00\% &                   38.68 &                   38.68\% \\
				\cmidrule(l){2-9}         &      \multirow{2}{*}{CNN}       &         \multicolumn{1}{c|}{Prune}          &                   42.44 &                   42.44\% &                    0.04 &                    0.04\% &                   42.40 &                   42.40\% \\
				&                                 &      \multicolumn{1}{c|}{Quantization}      &                   36.55 &                   36.55\% &                    0.04 &                    0.04\% &                   36.50 &                   36.50\% \\ \midrule
				\multirow{9}{*}{CIFAR-10}     &  \multirow{3}{*}{PlainNet-20}   &         \multicolumn{1}{c|}{Prune}          &                   50.40 &                   50.40\% &                   12.92 &                   12.92\% &                   37.48 &                   37.48\% \\
				&                                 &      \multicolumn{1}{c|}{Quantization}      &                   32.20 &                   32.20\% &                    8.16 &                    8.16\% &                   24.04 &                   24.04\% \\
				&                                 & \multicolumn{1}{c|}{Knowledge Distillation} &                   23.34 &                   23.34\% &                    6.77 &                    6.77\% &                   16.56 &                   16.56\% \\
				\cmidrule(l){2-9}         &   \multirow{3}{*}{ResNet-20}    &         \multicolumn{1}{c|}{Prune}          &                   49.32 &                   49.32\% &                   12.24 &                   12.24\% &                   37.08 &                   37.08\% \\
				&                                 &      \multicolumn{1}{c|}{Quantization}      &                   36.28 &                   36.28\% &                    6.44 &                    6.44\% &                   29.84 &                   29.84\% \\
				&                                 & \multicolumn{1}{c|}{Knowledge Distillation} &                   24.78 &                   24.78\% &                    6.38 &                    6.38\% &                   18.40 &                   18.40\% \\
				\cmidrule(l){2-9}         &     \multirow{3}{*}{VGG-16}     &         \multicolumn{1}{c|}{Prune}          &                   37.68 &                   37.68\% &                    5.12 &                    5.12\% &                   32.56 &                   32.56\% \\
				&                                 &      \multicolumn{1}{c|}{Quantization}      &                   25.64 &                   25.64\% &                    6.88 &                    6.88\% &                   18.76 &                   18.76\% \\
				&                                 &           Knowledge Distillation            &                   19.28 &                   19.28\% &                    4.78 &                    4.78\% &                   14.50 &                   14.50\% \\ \midrule
				\multirow{3}{*}{ImageNet}     &   \multirow{1}{*}{Inception}    &                Quantization                 &                   28.69 &                    28.69\% & 7.26 & 7.26\% &                   21.43 & 21.43\%  \\
				&   \multirow{1}{*}{ResNet-50}    &                Quantization                 &                   18.94 &  18.94\% &                  4.01 &   4.01\% &                 14.93 & 14.93\%  \\
				&  \multirow{1}{*}{ResNeXt-101}   &                Quantization                 &                   25.92 &   25.92\% &                  5.44 &    5.44\% &                   20.48 &               20.48\%   \\ \bottomrule
				\multicolumn{3}{c}{Average}                                            &                   32.73 &   32.73\% &                4.17 &  4.17\% &                   28.56 & 28.56\%  \\ \bottomrule
			\end{tabular}}
\end{table*}

\cref{tab:repair} shows the results.
On average, \repair repairs 32.73\% triggering inputs.
Although \repair induces 4.17\% new deviated behaviors,
overall \repair reduces the number of deviated behaviors by 30.16\%.
In the best case, the number of deviated behaviors is reduced by \RepairRate.
In conclusion,
it is feasible to repair the deviated behaviors using the triggering inputs found by \proj.
A promising feature work is to propose more advanced approaches to achieve this objective.

\begin{table*}[h]
    \centering
    {\caption{The effectiveness and efficiency of \proj  on the compressed model without \repair and with \repair.
            The numbers in parentheses are the ratios of time or queries spent by \proj on the compressed model repaired by \repair with respect the one spent by \proj on the model without \repair.
            The results are averaged across five runs using different random seeds.
        }
        \label{tab:retest}
\resizebox{\textwidth}{!}{\begin{tabular}{@{}cccc|rrr|rrr@{}}
                    \toprule
                    \multirow{4}{*}{Dataset}     & \multirow{4}{*}{Model}        & \multirow{4}{*}{Compression} &                  & \multicolumn{3}{c|}{Without \repair}                                                                                                                                       & \multicolumn{3}{c}{With \repair}                                                                                                                                  \\ \cmidrule(lr){5-7}\cmidrule(lr){8-10}
                    &                               &              &          Improvement                     & \multicolumn{1}{c}{\textbf{Average}} & \multicolumn{1}{c}{\textbf{Average}} & \multicolumn{1}{c|}{\multirow{2}{*}{\textbf{Average}}}  & \multicolumn{1}{c}{\textbf{Average}} & \multicolumn{1}{c}{\textbf{Average}}  & \multicolumn{1}{c}{\multirow{2}{*}{\textbf{Average}}} \\
                    &                               &                 &          Ratio                 & \multicolumn{1}{c}{\textbf{Success}} & \multicolumn{1}{c}{\textbf{Time}} & \multicolumn{1}{c|}{\multirow{2}{*}{\textbf{Query}}}  & \multicolumn{1}{c}{\textbf{Success}} & \multicolumn{1}{c}{\textbf{Time}}  & \multicolumn{1}{c}{\multirow{2}{*}{\textbf{Query}}} \\
                    &                               &                   &                          & \multicolumn{1}{c}{\textbf{Rate}} & \multicolumn{1}{c}{\textbf{(sec)}} &  & \multicolumn{1}{c}{\textbf{Rate}} & \multicolumn{1}{c}{\textbf{(sec)}}  &  \\ \midrule
                    \multirow{3}{*}{MNIST}       & LeNet-4      & \multicolumn{1}{c}{Prune}       &   48.48\%        & 100\%                                   &         0.056               & \multicolumn{1}{r|}{18.34} & 100\%                                   &          0.245 (4.83)& 55.67 (3.04)                                  \\
                    & LeNet-5                       & \multicolumn{1}{c}{Prune}       &   43.12\%        & 100\%         &      0.071                                     & \multicolumn{1}{r|}{    22.03 } & 100\%          &         0.169 (2.38)& 33.86 (1.54)                                                           \\
                    & \multirow{1}{*}{CNN}          & \multicolumn{1}{c}{Prune}          &    42.40\%    & 100\%                     &0.068               & \multicolumn{1}{r|}{22.51} & 100\%                                     &         0.399 (5.87)&  102.29       (4.54)                           \\ \midrule
                    \multirow{3}{*}{ImageNet}     &   \multirow{1}{*}{Inception}    &                Quantization                 &                   21.44\%&                    100\% & 1.266 & 21.44 &     100\%&              2.880 (2.27) &31.40 (1.46)  \\
                    &   \multirow{1}{*}{ResNet-50}    &                Quantization                 &                   18.94\% &  100\% &        0.819       &   19.24 &     100\%  &  2.959  (3.61)    &  39.45   (2.05)      \\
                    &  \multirow{1}{*}{ResNeXt-101}   &                Quantization                 &                   25.92\% &   100\% &           3.693       &   34.49  &         100\%  &       9.872 (2.67)     &      55.06 (1.60)            \\ \bottomrule
                \end{tabular}}
        }

    \end{table*}

We further leveraged \proj to test these models that are repaired by \repair.
Specifically, we selected the three models that have the highest improvement ratios to see if
these models that are relatively successfully repaired by \repair can decrease the effectiveness or efficiency \proj.
Meanwhile, we also selected the three models trained on ImageNet to
investigate the effects of \repair in large and complex models.
\cref{tab:retest} shows the effectiveness and efficiency of \proj when the compressed model
is not repaired by \repair and when the compressed model is repaired by \repair.
After repair, \proj can still achieve 100\% success rates in these six models.
However, the time spent by \proj to find each triggering input in the compressed model
repaired by \repair is 2.27x$\sim$5.87x as the one spent by \proj on the compressed models without repair.
The number of queries is also increased to 1.46x$\sim$4.54x as the one without repair.
As a proof of concept proposed by us,
\repair can effectively decrease the efficiency of \proj.
We will make the efforts to improve the effectiveness of \repair as our following work.

\answer{Answer to RQ5}{
    \repair reduces the deviated behaviors up to \RepairRate
    and decreases the efficiency of \proj.
    This result demonstrates the feasibility to repair
    the deviated behaviors using the triggering inputs found by \proj.
    We call for contributions from the community to propose more advanced approaches.
}

 \section{Discussion and Future Work}

\subsection{Demonstration of the Generalizability of \proj  on Other Domain}
\label{sec:audio}
Our study focuses on the compressed models for image classifications,
but our approach can also be applied to the compressed models in other domains
after proper adaptions, especially the mutation operators.
To demonstrate this, we applied \proj to the compressed models on
Speech-to-Text task.
Given an audio clip as input,
Speech-to-Text models aim to translate the audio into text.
We used the original models and compressed models provided
by Mozilla DeepSpeech~\cite{deepspeech}.\footnote{\url{https://github.com/mozilla/DeepSpeech}}
We selected Mozilla DeepSpeech since it is a
well-recognized open-source project (with more than 20,000 stars)
and it provides detailed documentation for us to deploy.
There are two pairs of original models and compressed model used in our evaluation.
Specifically,
the latest version of DeepSpeech, \ie, v0.9.3,
provides a pair of original model and compressed models
and the second latest version, v0.8.2, provides the second pair of models
(versions between these two versions provide the same models as v0.9.3).
In both version, the compressed models are quantized from the original models.

We adjusted \proj in two aspects to apply it in Speech-to-Text models.
First, we adopted the audio-specific mutation operators
since audio and images have different characteristics.
Specifically, we used the operators
TimeStretch, PitchShift, TimeShift, and Gain (volume adjustment)
provided by Audiomentations,\footnote{\url{https://github.com/iver56/audiomentations}}
a Python library to mutate audio.
Since these operators are also used in DeepSpeech for data
augmentation during model training,\footnote{\url{https://deepspeech.readthedocs.io/en/r0.9/TRAINING.html\#augmentation}}
we believe that these operators are regarded as representative mutations by developers.
Second, since the output of Speech-to-Text models is a sentence,
rather than a label in image classifications,
we also adjusted the methodology to compare the
outputs of original models and compressed models.
Specifically, in image classification models,
\proj compares the labels outputted by original models and compressed models,
while in Speech-to-Text, \proj compares the sentences word by word.
Given the same audio, if the original model and compressed model
output different sentences,
such as ``the character which your royal highness \emph{assumed} is imperfect harmony with your own''
vs ``the character which your royal highness \emph{summed} is imperfect harmony with your own'',
such an audio input is labeled as triggering input.
We also made the same adjustment to the baseline \df.
Three authors carefully reviewed the adjustment to avoid possible mistakes.

We randomly selected 500 audio inputs from the test set of Librispeech dataset~\cite{Librispeech}.\footnote{\url{https://www.openslr.org/12}}
According to the documentation,
Librispeech is used by Mozilla DeepSpeech in training and testing.
We used the same timeout as RQ1, \ie, 180 seconds.
The experiments were repeated five times using different random seeds.

\begin{table*}[h]
	\centering
	\caption{Effectiveness of \proj on on Speech-to-Text models.
		The results are averaged across five runs.}
	\label{tab:audio}
	\resizebox{\linewidth}{!}{
		\begin{tabular}{@{}ccc|rrr|rrr@{}}
				\toprule
				\multirow{4}{*}{Model}       &    \multirow{4}{*}{Version}     &        \multirow{4}{*}{Compression}         &                                                                \multicolumn{3}{c|}{\proj}                                                                 &                                                          \multicolumn{3}{c}{\df}                                                          \\
				\cmidrule(lr){4-6}\cmidrule(lr){7-9} &                               &                                             & \multicolumn{1}{c}{\textbf{Average}} &            \multicolumn{1}{c}{\textbf{Average}} &            \multicolumn{1}{c|}{\multirow{2}{*}{\textbf{Average}}} & \multicolumn{1}{c}{\textbf{Average}} &          \multicolumn{1}{c}{\textbf{Average}} & \multicolumn{1}{c}{\multirow{2}{*}{\textbf{Average}}} \\
				&                               &                                             & \multicolumn{1}{c}{\textbf{Success}} &               \multicolumn{1}{c}{\textbf{Time}} &              \multicolumn{1}{c|}{\multirow{2}{*}{\textbf{Query}}} & \multicolumn{1}{c}{\textbf{Success}} &             \multicolumn{1}{c}{\textbf{Time}} &   \multicolumn{1}{c}{\multirow{2}{*}{\textbf{Query}}} \\
				&                               &                                             &    \multicolumn{1}{c}{\textbf{Rate}} &              \multicolumn{1}{c}{\textbf{(sec)}} &                                                                   &    \multicolumn{1}{c}{\textbf{Rate}} &            \multicolumn{1}{c}{\textbf{(sec)}} &                                                       \\ \midrule
			 \multirow{2}{*}{DeepSpeech} &              v0.9.3&  Quantization &                100\% &                              5.740 &                                               8.42 & 95.6\% &169.365 & 223.04 \\
			&    v0.8.2  & Quantization  &                      100\% &                              4.812 &                                               5.88 &  95.5\% &168.402 &214.53 \\
			\bottomrule
		\end{tabular}
		}
\end{table*}

\cref{tab:audio} shows results.
The success rates of \proj are 100\% in five runs.
On average, it takes \proj
4.812$\sim$5.740s and 5.88$\sim$8.42 queries to find a triggering input.
By contrast, \df fails to find triggering inputs for around 4.5\% seed inputs
and it takes \df around 168 seconds and 214.53$\sim$223.04 queries to find one triggering input.
The time and queries spent by \proj is only 2.4\%$\sim$3.4\% and 2.7\%$\sim$3.8\% of the one required by \df, respectively.
This result demonstrates the effectiveness
and efficiency of \proj on Speech-to-Text tasks.

We also tried to fix the triggering inputs using \repair
but we were not able to achieve a reasonable result.
Our
conjecture is that repairing the models trained for Speech-to-Task
are much more complicated than the models trained for image classifications.
Specifically, for image classification models trained on ImageNet,
\repair is expected to output the label that is same as the label
outputted by the original model
from 1,000 candidate labels (since ImageNet has 1,000 image labels).
In contrast, for the Speech-to-Text task,
the output of models is a sentence that can have an arbitrary number of words
and there are around 977,000 unique English words in Librispeech.
To successfully repair the results outputted by compressed models,
\repair needs to not only select a correct set of words from these 977,000 words,
but also make sure these words are in the proper order
since the meaning of a sentence also depends on the order of words.
As a simple prototype proposed by us,
\repair is not able to handle such a complicated scenario.
We leave the improvement of \repair of large datasets like Librispeech for future work.

\subsection{Effect of Timeout}
In our evaluation, we used 180s as timeout for both \proj and \df.
To understand the effect of timeout on the effectiveness of \proj,
we conducted further experiments using smaller timeouts.
Specifically, we evaluated the success rate of \proj using 15s, 10s and 5s.
Our experiments covered all the pairs of models in RQ2 and used all the images from MNIST and CIFAR-10 test sets as seed inputs.

\proj achieves 100\% success rates for the 14 pairs of models out of \NumOriginalCompressedParis
pairs even using 5s as the timeout.
\cref{tab:timeout} shows the results of the remaining four pairs of models.
The success rates of \proj for these four pairs drop to different levels when the timeout is shortened.
The most significant decrease comes from the PlainNet-20 and its quantized model.
Specifically, its success rate drops to 76.93\% when the timeout is set to 15s.
The success rate drops further to 10.90\% with 5s timeout.
The success rate for VGG-16 and its compressed model also drops to 40.06\% with 5s timeout.
As for the other two pairs of models in \cref{tab:timeout}, their success rates slightly decrease
to 99.98\% and 89.12\% if 5s timeout is used, respectively.
In summary, \proj is effective for 16 out of \NumOriginalCompressedParis pairs of models
even a short timeout such as 10s is used.

An interesting observation from \cref{tab:timeout} is that
all the compressed models in ~\cref{tab:timeout} are compressed using 8-bit quantization.
A possible explanation is that the difference between an original model and its compressed model
induced by quantization is relatively smaller
than the difference induced by pruning and knowledge distillation.
Therefore, it takes a relatively long time for \proj to find the deviated behavior for quantized models.

\begin{table*}[h]
    \centering
    \caption{The success rates of \proj using different timeouts.
        The pairs of models that have 100\% success rate using 5s timeout
        are not included.
        The results are averaged across five runs.}
    \label{tab:timeout}
    \begin{tabular}{@{}ccc|rrr@{}}
        \toprule
        \multirow{2}{*}{\textbf{Dataset}} & \multirow{2}{*}{\textbf{Model}} & \multirow{2}{*}{\textbf{Compression}} &                            \multicolumn{3}{c}{Timeout}                             \\ \cmidrule(lr){4-6}
        &                                 &                                       & \multicolumn{1}{c}{15s} & \multicolumn{1}{c}{10s} & \multicolumn{1}{c}{5s} \\ \midrule
        CIFAR-10              &  \multicolumn{1}{l}{ResNet-20}  &          Quantization-8-bit           &                100\% &                100\% &                99.98\% \\ \midrule\midrule
        \multirow{4}{*}{CIFAR-10}     &  \multirow{1}{*}{PlainNet-20}   &             Quantization              &                 76.93\% &                 43.76\% &                10.90\% \\ \cmidrule(lr){2-6}
        &   \multirow{1}{*}{ResNet-20}    &             Quantization              &                 100\% &                  99.30\% &                89.12\% \\ \cmidrule(lr){2-6}
        &        \multirow{1}{*}{VGG-16}                         &   \multicolumn{1}{c|}{Quantization}   &                 95.87\% &                  79.30\% &                40.06\%                        \\ \bottomrule
    \end{tabular}\end{table*}

\subsection{Uniqueness of Triggering Inputs}
We carefully checked the triggering inputs found by \proj in~\cref{tab:results1}.
Specifically, we first represented each triggering input $x$ as a matrix $A_x$ with size $H\times W\times C$,
where $H$ and $W$ are the height and width of the image, respectively,
and $C$ refers to the number of channels of the image ($C=3$ in color images and $C=1$ in gray images).
Please note that the pixels in images are integers in the range $\left[0, 255\right]$,
and thus the elements in $A_x$ are also integer numbers in the range $\left[0, 255\right]$.
For each triggering input $x$, we check if there exists a triggering input $y$
such that the matrix $A_x$ is equal to the matrix $A_y$.
If such $y$ exists, the inputs $x$ and $y$ are labeled as duplicated triggering inputs.
Otherwise, $x$ is a unique triggering input.

Out of 105 experiments (\NumOriginalCompressedParis pairs of models $\times$ 5 runs),
77 experiments do not have any duplicated triggering inputs.
For the remaining 28 experiments,
on average, 99.04\% of the triggering inputs are unique to each other.
In other words,
the vast majority of the triggering inputs found by \proj are unique.

\subsection{Future Work}

A useful future work is to fix the deviated behaviors for compressed DNN models.
As we showed in~\cref{subsec:repair}, there is still a significant improvement space for
the performance of our repair prototype.
A promising research direction is to propose an effective and efficient
approach for this issue.

In~\cref{sec:audio}, we demonstrate the generalizability of \proj in Speech-to-Text tasks.
A promising direction is to apply \proj to other domains,
such as natural language process~\cite{bert} and object detection~\cite{fasterrcnn}.
To achieve this,
the mutation operators should be properly customized based on
domain-specific knowledge.
Meanwhile, the test oracle may be adjusted accordingly,
since the DNN models in other domains
may concern factors other than labels.
For example, in object detection, the location and boundary of the detected object are also important~\cite{map}.
Moreover, a sufficient number of compressed models and datasets from the AI community
are critical to comprehensively evaluate the new techniques in other domains.
We believe it is a fruitful working direction to explore.

Another potential follow-up direction is to leverage
\proj to directly test the DNN models deployed on the embedded or mobile platforms.
This may help the developers reveal the deviated behaviors induced
by the hardware or firmware of such platforms.
 \section{Threats to Validity}

\subsection{Internal Threats}
First, both \proj and \df have randomness at certain levels.
Such randomness may affect the evaluation results.
To alleviate this, all experiments are repeated five times using different random seeds and the average results are presented.
We found that the variance across these five runs are low
and the conclusions of our evaluation are consistent in each run,
\ie, \proj outperforms \df in both effectiveness and efficiency.
Therefore, we did not run the experiments more times.

Second, to comprehensively evaluate \proj using diverse compressed DNN models,
we construct the first benchmark containing \NumOriginalCompressedParis pairs of
original model and its compressed model.
Since there are no published pairs of the original model and its compressed model
for 15 out of \NumOriginalCompressedParis pairs,
we prepare them based on popular compression algorithms.
Specifically, we train the DNN models from scratch
and then compress them using popular model compression techniques.
Both processes may be affected by the randomness in deep learning at a certain level~\cite{variance}.
To mitigate this threat, for model training, we follow the practice from AI community
and train each model until the loss value is saturated.
We also compare their accuracy with the one reported by their original publications.
The accuracy of each model trained by us is close to its published accuracy.
In order to make sure that the model compressed by us are valid evaluation subjects,
we utilize an existing tool from Intel AI Lab~\cite{nzmora2019distiller}
and carefully follow the instructions.
The accuracy of each compressed model is close to that of its original model.
This suggests that our compression processes are reliable.

Third, it is possible that the triggering inputs found by \proj
do not comply with the real world data distribution.
To alleviate such a problem, \proj used the mutation operators from prior
work~\cite{deepxplore,diffchaser,deeptest, deepgauge,deephunter}
and these mutation operators are designed to simulate the
scenario that DNN models are likely to face in the real world.
For example, the mutation operator \textit{Random Pixel Change}
simulates effects of ``dirt on camera lens''~\cite{deepxplore}.
Gaussian Noise is one of the most frequently occurring noises in image signal~\cite{BONCELET2009143}.
Therefore, the triggering inputs found by \proj using these mutation operators
are highly likely to comply with the real-world inputs to be fed to DNN models in model deployment.

Lastly, since the implementation of \df shared by its authors only supports
image classification models, we carefully revised its source code
to support Speech-to-Text models in~\cref{sec:audio}.
It is possible that our revision might have mistakes and thus affects its effectiveness and efficiency.
To address this threat, three authors carefully reviewed the changes made by us
to avoid possible mistakes.

\subsection{External Threats}
We evaluate our approach using \NumOriginalCompressedParis compressed models.
The selection may not cover all compression techniques proposed by the communities.
To mitigate this, the models selected are representative as they are trained on two common datasets at different scales, and then compressed using popular model compression techniques.
Besides, the architectures of the selected models are diverse and include the  ones that are commonly used by existing studies~\cite{deepxplore, deeptest, emse21}.

\section{Related Work}

\subsection{DNN Model Testing}
DeepXplore~\cite{deepxplore} is the first technique targeted at testing DNN models.
It proposed neuron coverage, which measures the activation state of neurons, to guide the generation of test inputs.
DeepXplore is based on differential testing and it uses multiple models of a task to detect potential defects.
To alleviate the need of multiple models under test, DeepTest~\cite{deeptest} leverages metamorphic relations~\cite{chen1998metamorphic} that are expected to hold by a model as its test oracles.
Both DeepXplore and DeepTest perturb their test inputs based on the gradient of deep learning models.
TensorFuzz~\cite{tensorfuzz} and DeepHunter~\cite{deephunter} are whitebox fuzzing-based testing techniques.
They guide the input mutation by certain predefined coverage, instead of gradient, in order to trigger the unexpected behaviors of deep learning models, \eg numerical errors and classifications.
To assess the quality of DNN models,
    \textsc{DeepJanus}~\cite{deepjanus} proposes the notion of \textit{frontier of behaviors},
    \ie, pairs of inputs that have different predictions from the same DNN model.
    Given a DNN model under test, \textsc{DeepJanus}
    leverages a multi-objective evolutionary approach
    to find the frontier of behaviors.
    It further utilizes the model-based input representation to assure the realism of generated inputs.

Our approach, \proj, differs from these techniques in two ways.
First, \proj focuses on the deviated behaviors of compressed models,
while existing techniques target the normal models.
Second, the majority of existing testing techniques for DNN models are whitebox~\cite{deepxplore, deeptest,
	chen1998metamorphic, tensorfuzz, deephunter},
making use of the models' internal states, such as  gradients and neuron coverage, which are often unavailable for compressed models.
Therefore, these techniques are not applicable to testing compressed DNN models.
In contrast, our approach is specifically designed for compressed models and it does not require the internal information from the model under test.
The black-box testing approaches, \eg,~\textsc{DeepJanus}, with proper adaptations, are promising to be applied in finding deviated behaviors of compressed DNN models.
We will explore this direction in the future work.

Besides \df, there are also several recent studies specifically targeting on compressed models.
DiverGet~\cite{diverget} presents a search-based approach to assess quantization models
for hyperspectral images.
It proposes a set of domain-specific metamorphic relations to transform the hyperspectral images
and use them to mutate hyperspectral images.
BET~\cite{bet} is a testing method for convolutional neural network(CNN)s.
It splits a convolutional kernel into multiple zones of which the weights have the same positive or negative signs.
The insight is that the decisions of CNNs are likely to be affected by
continuous perturbations, \ie, the perturbations that have the same sign with each zone.
These two approaches are either specific to
the compression methods (quantization model in DiverGet) or types of DNN (CNN in BET),
while \proj is a general approach for diverse types of model architectures and compression methods.
We do not include their approach in our evaluation since their tools are not available.

\subsection{Empirical Study on DNN Model Deployment Issues}
Researchers have conducted several empirical studies to characterize the issues
in deploying DNN models, including compressed DNN models.
Guo~\etal found that the DNN models deployed in other platforms may exhibit different
behaviors from the original models~\cite{8952401}.
Hu~\etal conducted a deep analysis for quantization models~\cite{DBLP:journals/corr/abs-2204-04220}.
They found that retraining the compressed models with triggering inputs cannot effectively
reduce the behavioral difference between the original model and the compressed one.
Our approach, \proj, is a testing technique for compressed models,
with the aim to help developers address these issues in model deployment and dissemination.
Using the triggering inputs found by \proj, our prototype \repair is able to repair up to \RepairRate deviated behaviors.

\subsection{Differential Testing}
\proj aims to find deviated behavior between two DNN models.
Related works also include those applying differential testing to detect inconsistencies across two pieces of traditional software.
McKeeman~\cite{differential_testing} originally proposed differential testing in 1998 to expose bugs in software systems using test cases that result in inconsistent execution results in multiple comparable systems.
Le~\etal~\cite{EMI} introduced EMI, which applies differential testing on compilers using semantically equivalent programs.
Inconsistent execution outputs of compiled programs may indicate defects in compilers.
Further, differential testing is also applied in JVM implementations~\cite{JVM} using mutated Java bytecode~\cite{classming}.

The objective of \proj is similar to differential testing.
Rather than two pieces of code, the systems under test for \proj are DNN models and their compressed ones.

\subsection{Differential Verification of DNN models}
ReluDiff~\cite{reludiff} and its following work~\cite{neurodiff}  share certain similar objectives with our approach although it is not a testing technique.
It leverages the structural and behavioral similarities of the two closely related networks in parallel, to verify whether the output difference of the two models are within the specification.
In the evaluation, they use the pairs of compressed model and the original model as subjects.

Our work differs from ReluDiff in two ways.
First, ReluDiff can only be used in forward neural networks with relu activation function for both compressed and original models.
This limits its application scenarios.
Sophisticated DNN models usually contain convolutional layers and recurrent layers.
The advantage of \proj is that it makes no assumption on the model architecture, making it applicable to a wide range of application scenarios.
Second, ReluDiff needs to know the architectures and weights of DNN models for verification, while \proj works for black-box models.

\section{Conclusion}
We proposed \proj, a novel, effective input generation method to find deviated behaviors between an original DNN model and its compressed model.
Specifically, \proj leverages the MH algorithm in the selection of a mutation operator at each iteration to successively mutate a given seed input.
\proj incorporates a novel fitness function to determine whether to use a mutated input in subsequent iterations.
The results show that \proj outperforms prior work in terms of both effectiveness and efficiency.
\proj constantly achieves 100\% success rate
but uses significantly less amount of time and queries than the state of the art.
We also explored the possibility to repair such deviated behaviors using the triggering inputs found by \proj.
Our prototype \repair can repair up to \RepairRate deviated behaviors
and decrease the effectiveness of \proj on the repaired models.

\begin{acks}
We would like to thank the editor and reviewers of TOSEM for their constructive comments on this study.

The authors at the Hong Kong University of Science and Technology
are supported by
National Natural Science Foundation of China (Grant No: 61932021),
Hong Kong RGC/GRF (Grant No: 16207120),
Hong Kong RGC/RIF (Grant No: R5034-18),
Hong Kong ITF (Grant No: MHP/055/19),
Hong Kong PhD Fellowship Scheme,
HKUST RedBird Academic Excellence Award,
and the MSRA Collaborative Research Grant.
The authors at University of Waterloo are supported by
Cisco Research Gift,
Natural Sciences and Engineering Research Council of Canada (NSERC) through the
Discovery Grant,
and CFI-JELF Project \#40736.

\end{acks}
\bibliographystyle{ACM-Reference-Format}
\bibliography{tosem}
\end{document}